\documentclass[12pt, preprint]{aastex}

\slugcomment{Accepted for Publication in {\it The Astrophysical Journal}}

\shorttitle{Leptonic and Hadronic Modeling of Fermi-Detected Blazars}
\shortauthors{M. B\"ottcher et al.}

\begin{document}

\title{Leptonic and Hadronic Modeling of Fermi-Detected Blazars}

\author{M. B\"ottcher\altaffilmark{1,2}, A. Reimer\altaffilmark{3},
K. Sweeney\altaffilmark{2} and A. Prakash\altaffilmark{2}}

\altaffiltext{1}{Centre for Space Research, North-West University, Potchefstroom,
2531, South Africa}

\altaffiltext{2}{Astrophysical Institute, Department of Physics and Astronomy, \\
Ohio University, Athens, OH 45701, USA}

\altaffiltext{3}{Institut f\"ur Theoretische Physik, 
Universit\"at Innsbruck, Technikerstra\ss e 25, A-6020 Innsbruck, 
Austria}

\begin{abstract}
We describe new implementations of leptonic and hadronic models for
the broadband emission from relativistic jets in AGN in a temporary
steady state. For the leptonic model, a temporary equilibrium between
particle injection/acceleration, radiative cooling, and escape from
a spherical emission region is evaluated, and the self-consistent
radiative output is calculated. For 
the hadronic model, a temporary equilibrium between particle 
injection/acceleration, radiative and adiabatic cooling, and
escape is evaluated for both primary electrons and protons.
A new, semi-analytical
method to evaluate the radiative output from cascades initiated
by internal $\gamma\gamma$ pair production is presented. We use
our codes to fit snap-shot spectral energy distributions of a
representative set of {\it Fermi}-LAT detected blazars. 
We find that the leptonic model provides
acceptable fits to the SEDs of almost all blazars with 
parameters close to equipartition between the magnetic field
and the relativistic electron population. However, the hard
$\gamma$-ray spectrum of AO~0235+164, in contrast to the very
steep IR-optical-UV continuum, poses a severe problem for the
leptonic model. If charge
neutrality in leptonic models is provided by cold protons, 
the kinetic energy carried by the jet should
be dominated by protons. We find satisfactory representations of 
the snapshot SEDs of most blazars in our sample with the hadronic 
model presented here. However, in the case of two quasars the 
characteristic break at a few GeV energies can not be well modelled.
All of our hadronic model fits require powers in relativistic protons 
in the range $L_p \sim 10^{47}$ -- $10^{49}$~erg~s$^{-1}$.

\end{abstract}

\section{Introduction}

Blazars are a class of radio-loud active galactic nuclei (AGNs) 
comprised of Flat-Spectrum Radio Quasars (FSRQs) and BL~Lac objects. 
Their spectral energy distributions (SEDs) are characterized by 
non-thermal continuum spectra with a broad low-frequency component
in the radio -- UV or X-ray frequency range and a high-frequency
component from X-rays to $\gamma$-rays, and they often exhibit
substantial variability across the electromagnetic spectrum,
in extreme cases on time scales down to just a few minutes. 
Blazars are sub-divided based on the location of their synchrotron
peak: Low-Synchrotron-Peaked (LSP) blazars, consisting of FSRQs and
Low-frequency-peaked BL Lac objects (LBLs) have $\nu_{\rm sy}
< 10^{14}$~Hz; Intermediate-Synchrotron-Peaked (ISP) blazars 
(which are exclusively BL~Lac objects and hence also termed
IBLs for Intermediate BL Lac Objects) have $10^{14}$~Hz~$\le 
\nu_{\rm sy} \le 10^{15}$~Hz; High-Synchrotron-Peaked (HSP)
blazars (also exclusively BL Lac objects, termed HBLs for
High-peaked BL Lac objects) have $\nu_s > 10^{15}$~Hz.

The extreme inferred isotropic luminosities, combined with the
rapid high-energy variability, provide convincing evidence that
the high-energy emission from blazars originates in relativistic
jets closely aligned with our line of sight \citep[for a review
of those arguments, see, e.g.,][]{schlickeiser96}. The simplest
(and often sufficient) assumption is that the emission is 
produced in an approximately spherical region, propagating
along the jet with a speed $\beta_{\Gamma} c$, corresponding to
a bulk Lorentz factor $\Gamma$. If the jet forms an angle 
$\theta_{\rm obs}$ with respect to our line of sight, this 
results in Doppler boosting characterized by the Doppler factor 
$D = \left( \Gamma \, [1 - \beta_{\Gamma} \cos\theta_{\rm obs}] 
\right)^{-1}$. The observed bolometric flux will be enhanced
by a factor $D^4$, while photon energies are blue-shifted by
a factor $D$, and variability time scales will be shortened
by a factor $D^{-1}$ \cite[for a review of relativistic effects 
in the jets of AGN, see][]{boettcher12}.

It is generally accepted that the low-frequency (radio through
UV or X-ray) emission from blazars is synchrotron emission from
relativistic electrons in the jet. For the origin of the high-energy
(X-ray through $\gamma$-ray) emission, two fundamentally different
approaches have been proposed, generally referred to as leptonic
and hadronic models. 

In leptonic models, the radiative output throughout the electromagnetic
spectrum is assumed to be dominated by leptons (electrons and possibly
positrons), while any protons that are likely present in the outflow,
are not accelerated to sufficiently high energies to contribute 
significantly to the radiative output. The high-energy emission is
then most plausibly explained by Compton scattering of low-energy
photons by the same electrons producing the synchrotron emission
at lower frequencies \citep{mcg92,bm96,ds93,sikora94,blaz00}. 
In hadronic models, both primary electrons and protons are
accelerated to ultrarelativistic energies, with protons exceeding
the threshold for p$\gamma$ photo-pion production on the soft
photon field in the emission region. While the low-frequency
emission is still dominated by synchrotron emission from primary
electrons, the high-energy emission is dominated by proton 
synchrotron emission, $\pi^o$ decay photons, synchrotron and 
Compton emission from secondary decay products of charged pions,
and the output from pair cascades initiated by these high-energy
emissions intrinsically absorbed by $\gamma\gamma$ pair production
\citep{mb92,aharonian00,mp01,muecke03}. For a general overview
of the features of both leptonic and hadronic blazar models,
see \cite{boettcher10}.

Steady-state leptonic models have met with great success in 
modeling the spectral energy distributions (SEDs) of all classes 
of blazars \citep[e.g.,][]{ghisellini98,cg08}, including substantial 
samples of {\it Fermi}-detected blazars \citep[e.g.,][]{ghisellini10}.
Much progress has also been made to explore the variability features 
predicted in time-dependent implementations of leptonic models 
\citep[e.g.,][]{mk97,kusunose00,lk00,bc02,sokolov04,graff08,bd10,jb11}.
However, the very fast (time scales of a few minutes) variability 
of some TeV blazars \citep{albert07,aharonian07} poses severe
problems for single-zone models of blazars due to the required 
extremely high bulk Lorentz factors \citep{begelman08}. Even without 
considering variability, the SED of the quasar 3C~279, detected in 
VHE $\gamma$-rays by MAGIC \citep{albert08}, poses problems for 
one-zone leptonic models, and is more easily represented by a hadronic 
model \citep{brm09}. In order to remedy some of the problems, several 
variations of multi-zone models have been proposed, including the 
spine-sheath model of \cite{tg08} or the decelerating-jet model of 
\cite{gk03}, as well as internal-shock models 
\citep[e.g.,][]{mg85,spada01,sokolov04,graff08,jb11,bd10}.

The goal of this work is to provide tools for the modeling of blazar 
spectra including data from {\it Fermi}-LAT, which have been accumulated 
over one year of {\it Fermi} operations. Any short-term variability in
the SEDs of the blazars under consideration, has therefore been averaged
out over this integration time. Therefore, for the purpose of this study,
we will use time-independent models, which represent an appropriate time 
average of the rapidly variable broadband emission. 

The accurate evaluation of the radiative output of hadronic
models is best achieved by Monte-Carlo simulations 
\citep[e.g.,][]{muecke00} due to the complicated 
energy dependence of the cross sections (in particular, for
p$\gamma$ interactions) involved. As those are quite time-consuming,
hadronic models have so far received much less attention in the
literature than leptonic ones, and in particular, time dependent
implementations of hadronic models are still in their early
development stages. A comparison between the results of fitting 
a large, statistically representative sample of blazars with both 
leptonic and hadronic models, considered to comparable degree of 
detail, has therefore never been performed. This is the main purpose
of the code developments and modeling efforts presented in this paper.

In \S \ref{lepton} we describe recent developments to our
existing leptonic radiation transfer code, in particular
its application to quasi-stationary situations, and the
implementation of arbitrary external radiation fields as
sources for Compton scattering. In \S \ref{hadron} we
will describe a new, semi-analytical implementation of 
the stationary hadronic model
and test it against results of detailed Monte-Carlo simulations
based on the SOPHIA code \citep{muecke00}. We present the
results of our comparative leptonic and hadronic modeling 
of the SEDs of a representative sample of {\it Fermi}-LAT
detected blazars of different sub-classes in \S \ref{modeling}. 
We summarize and discuss our results in \S \ref{summary}. 
Throughout our paper, we convert redshifts to luminosity 
distances using a $\Lambda$CDM cosmology with 
$H_0 = 70$~km~s$^{-1}$~Mpc$^{-1}$, $\Omega_m = 0.3$ and 
$\Omega_{\Lambda} = 0.7$.

\section{\label{lepton}Stationary leptonic blazar model}

The leptonic jet radiation transfer model used for this work
is based on the work of \cite{bc02} \citep[see also][]{bms97,bb00}.
It is a homogeneous one-zone model in which a population of 
ultrarelativistic electrons (or positrons) is injected with a 
power-law distribution $Q_e (\gamma_e) = Q_0 \, \gamma_e^{-q_e} \, 
H(\gamma_e; \gamma_{e,1}, \gamma_{e,2})$ into a spherical emission region
of co-moving radius $R$. Here, $H(x; a, b)$ is the Heaviside
function defined as $H = 1$ if $a \le x \le b$ and $H = 0$
otherwise. The emission region moves with constant relativistic
speed $\beta_{\Gamma} c$ (corresponding to bulk Lorentz factor
$\Gamma$) along the jet. 

The electron distribution cools due to synchrotron and Compton 
emission. Synchrotron emission is determined through a tangled 
magnetic field of co-moving strength $B$. For Compton scattering, 
the synchrotron radiation field (SSC = synchrotron self-Compton) 
and various external radiation fields are taken into account:
(a) Direct accretion disk emission \citep[see][for details]{bms97}, 
(b) accretion disk emission re-processed by the Broad Line Region
\citep[BLR; see][for details]{bb00}, (c) an isotropic (in the rest 
frame of the AGN) external radiation field of arbitrary spectral shape. 
For the latter, we have developed a pre-processing routine which
produces a spectrum file of the external radiation field in the
AGN rest frame. The blazar code applies the proper relativistic
Lorentz transformations into the blob rest frame to use this
radiation field as sources for external Compton scattering. Such
a description is appropriate for radiation fields for which the
angular characteristics of the radiation field in the co-moving
frame are dominated by relativistic aberration rather than any 
intrinsic anisotropy. Quantitatively, the characteristic scale 
of angular variations $\Delta\theta_{\rm ext}$ of the external 
radiation field should be $\Delta\theta_{\rm ext} \gg 1/\Gamma$.
This is typically the case for (1) (line-dominated) emission from
the BLR as long as the emission region is located within the
BLR; (2) infrared emission from a large-scale dusty torus around
the central engine; (3) the Cosmic Microwave Background. The
external Compton emissivity is evaluated using the head-on approximation
for the Compton cross section \citep[based on an angle integration of
the full Klein-Nishina cross section; see, e.g.,][]{dm09}. 
The electron cooling rates due to Compton scattering are calculated
using the full Klein-Nishina cross section, adopting the analytical 
solution of \cite{bms97}. 
Particles may escape the emission region on a time scale $t_{\rm esc} 
\equiv \eta_{\rm esc} \, R / c$, which we parameterize with an escape 
time scale parameter $\eta_{\rm esc} \ge 1$. 

In order to fit SEDs of blazars in the absence of detailed spectral 
variability information, it is appropriate (and computationally more 
efficient than detailed time-dependent radiation transfer simulations) 
to apply a stationary radiation transfer model. For the stationary 
model used in this work, our code finds an equilibrium between the
relativistic particle injection mentioned above, radiative cooling,
and escape. For this purpose, a critical Lorentz factor $\gamma_c$
is determined, for which the radiative cooling time scale equals
the escape time scale, $\gamma_c / \vert\dot\gamma (\gamma_c) \vert 
= t_{\rm esc}$. The shape of the resulting equilibrium particle 
distribution will then depend on whether $\gamma_c < \gamma_{e,1}$
(the fast cooling regime) or $\gamma_c > \gamma_{e,1}$ (the slow cooling
regime). In the fast cooling regime,

\begin{equation}
n_e^{\rm f.c.} (\gamma_e) = n_0 \cases{ 
\left( {\gamma_e \over \gamma_{e,1}} \right)^{-2} & for $\gamma_c \le \gamma_e \le \gamma_{e,1}$ \cr\cr
\left( {\gamma_e \over \gamma_{e,1}} \right)^{-(q_e + 1)} & for $\gamma_{e,1} < \gamma_e < \gamma_{e,2}$ \cr\cr
0 & else \cr }
\label{Nefc}
\end{equation}

while in the slow cooling regime,

\begin{equation}
n_e^{\rm s.c.} (\gamma_e) = n_0 \cases{ 
\left( {\gamma_e \over \gamma_c} \right)^{-q_e} & for $\gamma_{e,1} \le \gamma_e \le \gamma_c$ \cr\cr
\left( {\gamma_e \over \gamma_c} \right)^{-(q_e + 1)} & for $\gamma_c < \gamma_e < \gamma_{e,2}$ \cr\cr
0 & else \cr }
\label{Nesc}
\end{equation}

Since the radiative cooling depends on the self-consistent radiation
field, including synchrotron emission from the present relativistic
electron population, an iterative scheme is applied: The code starts
out with an equilibrium based on only synchrotron and external-Compton
cooling. This distribution is then used to calculate the synchrotron
radiation energy density, which is added to the total radiation energy
density to re-calculate $\gamma_c$ and the resulting equilibrium particle
distribution. The process is iterated until convergence is achieved.

The code then evaluates the resulting kinetic power in relativistic
electrons in the AGN frame,

\begin{equation}
L_e = \pi \, R^2 \, \Gamma^2 \, \beta_{\Gamma} \, c \, m_e c^2 \, 
\int\limits_1^{\infty} d\gamma_e \; n_e (\gamma_e) \, \gamma_e
\label{Le}
\end{equation}
and compares it to the power carried in the magnetic field (Poynting
flux),

\begin{equation}
L_B = \pi \, R^2 \, \Gamma^2 \, \beta_{\Gamma} \, c \, {B^2 \over 8 \pi}
\label{LB}
\end{equation}
We define the equipartition parameter as the ratio of the two, i.e.,
$\epsilon_{Be} \equiv L_B / L_e$.
We also evaluate the kinetic luminosity
in cold protons that are expected to be present if charge neutrality 
is provided by one cold proton per electron (i.e., no pairs):

\begin{equation}
L_p = \pi \, R^2 \, \Gamma^2 \, \beta_{\Gamma} \, c \, n_e \, m_p c^2 
\label{Lp}
\end{equation}
The presence of pairs in addition to an electron-proton plasma in the
jet will lower the total kinetic luminosity of the jet, $L_{\rm kin}
\equiv L_e + L_p$. 

\begin{figure}[ht]
\vskip 1cm
\centering
\includegraphics[width=12cm]{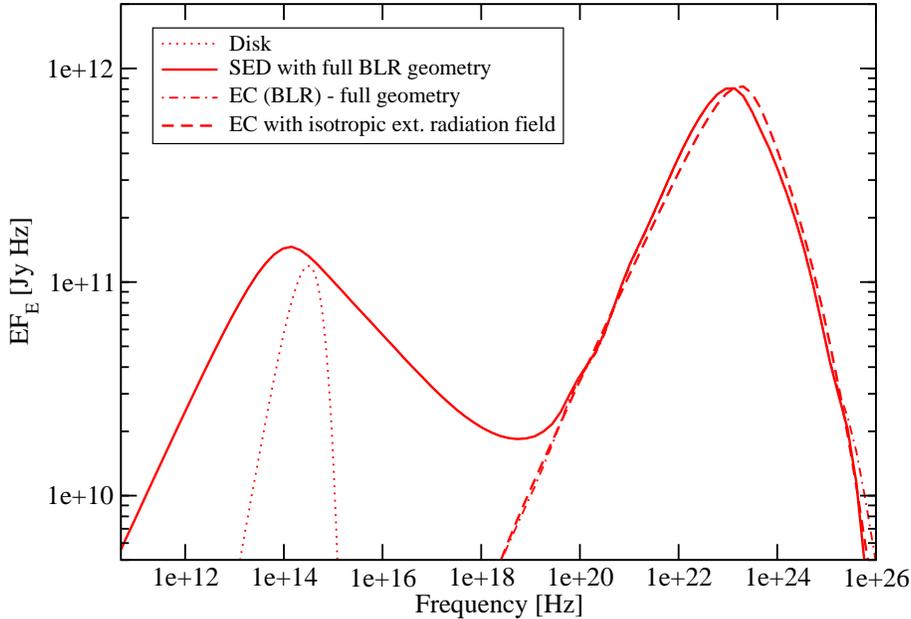}
\caption{\label{ECtest}Generic blazar SEDs with EC(BLR) calculated with
the full BLR geometry (solid SED; EC[BLR] component: dot-dashed line).
The dotted line shows the disk spectrum. The dashed line shows the EC
component calculated using an isotropic, thermal external radiation field 
with a temperature producing the same peak frequency as the disk, and the 
same radiation energy density as resulting from BLR reprocessing of the 
disk emission. }
\end{figure}

The solid line in Figure \ref{ECtest} shows a generic model blazar SED
with an EC component calculated with the full BLR re-processing calculation 
described in \cite{bb00} (dot-dashed line) for a blazar located at a redshift 
of $z = 0.3$. For this calculation, we assumed a Shakura-Sunyaev accretion 
disk with a luminosity of $L_D = 10^{45}$~erg~s$^{-1}$ with a multi-color 
blackbody spectrum (the dotted line in Figure \ref{ECtest}) peaking at $\nu_{D} 
\approx 3.9 \times 10^{14}$~Hz in the AGN rest frame. A spherical BLR with
re-processing fraction $\eta_{\rm BLR} = 0.04$ is placed at a distance 
$R_{\rm BLR} = 0.2$~pc from the accretion disk. The average radiation 
energy density inside the BLR is commonly estimated as

\begin{equation}
u_{\rm BLR} = {L_D \, \eta_{\rm BLR} \over 4 \pi \, R_{\rm BLR}^2 \, c} 
\approx 2.9 \times 10^{-4} \; {\rm erg \; cm}^{-3}
\label{uBLR}
\end{equation}
In an emission region located close to the inner edge of the BLR, an 
electron distribution is injected with a power-law with index $q = 2.5$ 
between $\gamma_1 = 10^3$ and $\gamma_2 = 5 \times 10^5$. The equilibrium
electron population is in the fast cooling regime.

For comparison, the dashed line in Figure \ref{ECtest} shows the
(much faster) calculation of the EC component using an isotropic,
thermal radiation field with a blackbody temperature $T = 5000$~K, 
reproducing a peak energy close to the one of accretion disk, and 
with an energy density of $u_{\rm ext} = 2.5 \times 10^{-4}$~erg~cm$^{-3}$.
The figure shows that the isotropic radiation field approximation with
these parameters provides an excellent description of the EC component. This 
indicates that the approximation of Eq. (\ref{uBLR}) slightly over-estimates 
the energy density of the radiation field. In the comprehensive modeling 
effort described in \S \ref{modeling}, we represent all BLR components 
as isotropic radiation fields, keeping in mind that a simple translation 
of disk and BLR parameters to the external radiation field energy density 
of Eq. (\ref{uBLR}) needs to be taken with caution. 

The model described above has already been used successfully to model 
a number of blazar SEDs. In particular, it has been used to interpret
the SEDs of blazars detected in very high energies (VHE, $E > 100$~GeV)
by the Very Energetic Radiation Imaging Telescope Array System (VERITAS),
e.g., W Comae \citep{WComa,WComb}, 1ES~0806+524 \citep{0806}, 
PKS~1424+240 \citep{1424}, RGB~J0710+591 \citep{0710}, and 3C~66A 
\citep{3C66A}. It has also been used to model the SED of the high-redshift
FSRQ PKS~0528+134 in quiescence \citep{palma11}.

\section{\label{hadron}Stationary hadronic blazar model}

In order for protons to contribute significantly to the radiative
output in relativistic jets of blazars either through proton
synchrotron emission \citep{aharonian00,mp00} or photo-pion production
\citep{mb92}, they need to be accelerated to proton energies of 
typically $E_p \equiv E_{19} \, 10^{19}$~eV with $E_{19} \gtrsim 1$. 
Requiring that the Larmor radius of these protons be smaller than the 
size of the emission region, $R \equiv 10^{15} \, R_{15}$~cm, one
needs magnetic fields of $B \ge 30 \, E_{19} \, R_{15}^{-1}$~G.
The energy density of such large magnetic fields is likely to
dominate over the energy density of external radiation fields
in typical AGN environments, so that the synchrotron radiation
field is likely to be the dominant target photon field for
photo-pion production by ultrarelativistic protons. We will
therefore restrict our considerations in this paper to the
synchrotron-proton blazar model as discussed in detail in
\cite{mp01} and \cite{muecke03}.

For a comparison between leptonic and hadronic blazar models
on the basis of a sufficiently large sample, we need a 
hadronic model implementation that allows for the efficient
scanning through parameter space within a reasonable time frame. 
At the same time, the model needs to treat hadronic processes to
a comparable degree of detail as the leptonic processes are being
considered in the leptonic model described in the previous section.
The most accurate and reliable method to evaluate the radiative 
output from photo-pair and photo-pion production and the cascades 
initiated by $\gamma\gamma$ absorption of ultra-high-energy (UHE; 
$E \gg 1$~TeV) photons produced through $\pi^0$ decay and synchrotron 
emission of pions and their decay products, is through Monte-Carlo
simulations \citep[e.g.,][]{muecke00}. However, such simulations 
are quite time consuming, and a comprehensive modeling effort of 
a large sample of SEDs is currently infeasible with this method.

In order to avoid time-consuming Monte-Carlo simulations, \cite{ka08}
have developed analytical expressions for the final decay products
(electrons, positrons, photons, neutrinos) from photo-pion production
for isotropically distributed, mono-energetic photons and protons. 
These can be integrated over
any proton and photon distribution to obtain the final production
spectra of electrons, positrons, photons, and neutrinos. However,
the $\pi^0$ decay photons as well as synchrotron photons from these
first-generation electrons and positrons emerge in the UHE regime, 
at which the dense radiation fields in the emission regions of blazars 
are highly opaque to $\gamma\gamma$ pair production. It is therefore 
essential to include the effects of UHE $\gamma$-ray induced pair 
cascades in hadronic models. Again, time-consuming Monte-Carlo/numerical
simulations constitute the standard method for the evaluation of 
such cascades.

\subsection{\label{cascades}Synchrotron-Supported Pair Cascades}

For the purpose of an efficient calculation of the cascades, we have
developed a semi-analytical method that does not involve Monte-Carlo
simulations. Let us assume that the injection rates of first-generation
high-energy $\gamma$-rays, $\dot N_{\epsilon}^0$, and pairs, $Q_e (\gamma)$,
are known, e.g., from the analytical approximations of \cite{ka08}. 
Throughout this exposition, we use the dimensionless photon energy 
$\epsilon \equiv h \nu / m_e c^2$. In the case of linear cascades the
optical depth for $\gamma\gamma$ absorption, $\tau_{\gamma\gamma} 
(\epsilon)$ can be pre-calculated from the low-energy radiation field. 
Under these conditions, the spectrum of escaping (observable) photons 
can be calculated as

\begin{equation}
\dot N_{\epsilon}^{\rm esc} = \dot N_{\epsilon}^{\rm em} \, \left( 
{1 - e^{-\tau_{\gamma\gamma} [\epsilon]} \over \tau_{\gamma\gamma} [\epsilon]}
\right)
\label{Ndotescape}
\end{equation}
where
$\dot N_{\epsilon}^{\rm em}$ has contributions from the first-generation 
high-energy photon spectrum and synchrotron emission from secondaries,
$\dot N_{\epsilon}^{\rm em} = \dot N_{\epsilon}^0 + \dot N_{\epsilon}^{\rm sy}$.
We use a synchrotron emissivity function for a single electron of the form 
$j_{\nu} \propto \nu^{1/3} \, e^{-\epsilon/\epsilon_0}$ with $\epsilon_0 = 
b \, \gamma^2$, where $b \equiv B / B_{\rm crit}$ and $B_{\rm crit} = 4.4 
\times 10^{13}$~G. This yields

\begin{equation}
\dot N_{\epsilon}^{\rm sy} = A_0 \, \epsilon^{-2/3} \, \int\limits_1^{\infty}
d\gamma \, N_e (\gamma) \, \gamma^{-2/3} \, e^{-\epsilon/(b \gamma^2)}
\label{Ndotsy}
\end{equation}
with the normalization

\begin{equation}
A_0  = {c \, \sigma_T \, B^2 \over 6 \pi \, m_e c^2 \, \Gamma(4/3) \, b^{4/3}}
\label{Nsynorm}
\end{equation}

The electron distribution, $N_e (\gamma)$ will be the solution to the isotropic
Fokker-Planck equation in equilibrium ($\partial \langle . \rangle / \partial t
= 0$):

\begin{equation}
{\partial \over \partial \gamma} \left( \dot\gamma \, N_e [\gamma] \right)
= Q_e (\gamma) + \dot N_e^{\gamma\gamma} (\gamma) + \dot N_e (\gamma)^{\rm esc}.
\label{fp}
\end{equation}
In the case under consideration here, the electron energy losses will be 
dominated by synchrotron losses, i.e.,

\begin{equation}
\dot\gamma = - {c \, \sigma_T \, B^2 \over 6 \pi \, m_e c^2} \, \gamma^2
\equiv - \nu_0 \gamma^2.
\label{gammadotsy}
\end{equation}
We represent the escape term in Eq. (\ref{fp}) through an energy-independent
escape time scale $t_{\rm esc} = \eta_{\rm esc} \, R / c$, so that 
$\dot N_e (\gamma)^{\rm esc} = - N_e (\gamma) / t_{\rm esc}$.
$\dot N_e^{\gamma\gamma} (\gamma)$ in Eq. \ref{fp} is the rate of particle 
injection due to
$\gamma\gamma$ absorption, to be evaluated self-consistently 
with the radiation
field. In the $\gamma\gamma$ absorption of a high-energy 
photon of energy
$\epsilon$, one of the produced particles will assume the 
major fraction, $f_{\gamma}$ of the photon energy. From comparison with
Monte-Carlo simulations, we find that $f_{\gamma} = 0.9$ yields good
agreement with numerical solutions to the cascade problem. 
Hence, an electron/positron 
pair with energies $\gamma_1 = f_{\gamma} \, \epsilon$ and $\gamma_2 = 
(1 - f_{\gamma}) \, \epsilon$ is produced. Furthermore realizing that every 
photon not escaping (according to Eq.
\ref{Ndotescape}) will produce an 
electron/positron pair, we can write the pair production rate as

\begin{equation}
\dot N_e^{\gamma\gamma} (\gamma) = f_{\rm abs} (\epsilon_1) \, 
\left( \dot N_{\epsilon_1}^0 + \dot N_{\epsilon_1}^{\rm sy} \right)
+ f_{\rm abs} (\epsilon_2) \, \left( \dot N_{\epsilon_2}^0 
+ \dot N_{\epsilon_2}^{\rm sy} \right)
\label{Ndotgamma}
\end{equation}
where $\epsilon_1 = \gamma/f_{\gamma}$, $\epsilon_2 = \gamma/(1 - f_{\gamma})$
and

\begin{equation}
f_{\rm abs} (\epsilon) \equiv 1 - {1 - e^{-\tau_{\gamma\gamma} (\epsilon)}
\over \tau_{\gamma\gamma} (\epsilon)}
\label{fabs}
\end{equation}
With this approximation, we find an implicit solution to Equation \ref{fp}:

\begin{equation}
N_e (\gamma) = {1 \over \nu_0 \gamma^2} \int\limits_{\gamma}^{\infty} d\tilde\gamma
\left\lbrace Q_e (\tilde\gamma) + \dot N_e^{\gamma\gamma} (\tilde\gamma) 
- {N_e (\tilde\gamma) \over t_{\rm esc}}
\right\rbrace 
\label{Nsolution}
\end{equation}
The solution (\ref{Nsolution}) is implicit in the sense that the particle
spectrum $N_e (\gamma)$ occurs on both sides of the equation as $\dot 
N_e^{\gamma\gamma}$ depends on the synchrotron emissivity calculated through
Eq. \ref{Ndotsy}, which requires knowledge of $N_e (\tilde\gamma)$, where
pairs at energies of $\tilde\gamma_1 = \sqrt{\gamma / (f_{\gamma} b)}$ 
and $\tilde\gamma_2 = \sqrt{\gamma / ([1 - f_{\gamma}] \, b)}$ provide 
the majority of the radiative output relevant for pair production at 
energy $\gamma$. However, for practical applications, we may use the 
fact that generally, $\gamma$, the argument on the l.h.s., 
is much smaller than $\tilde\gamma_{1,2}$. 
Therefore, Eq. \ref{Nsolution} may be evaluated progressively, starting at the 
highest pair energies for which $Q_0 (\gamma) \ne 0$ or $\dot N_{\epsilon_{1,2}}^0
\ne 0$, and then using the solution for $N_e (\gamma)$ for large $\gamma$
as one progresses towards lower values of $\gamma$.

\begin{figure}[ht]
\vskip 1cm
\centering
\includegraphics[width=12cm]{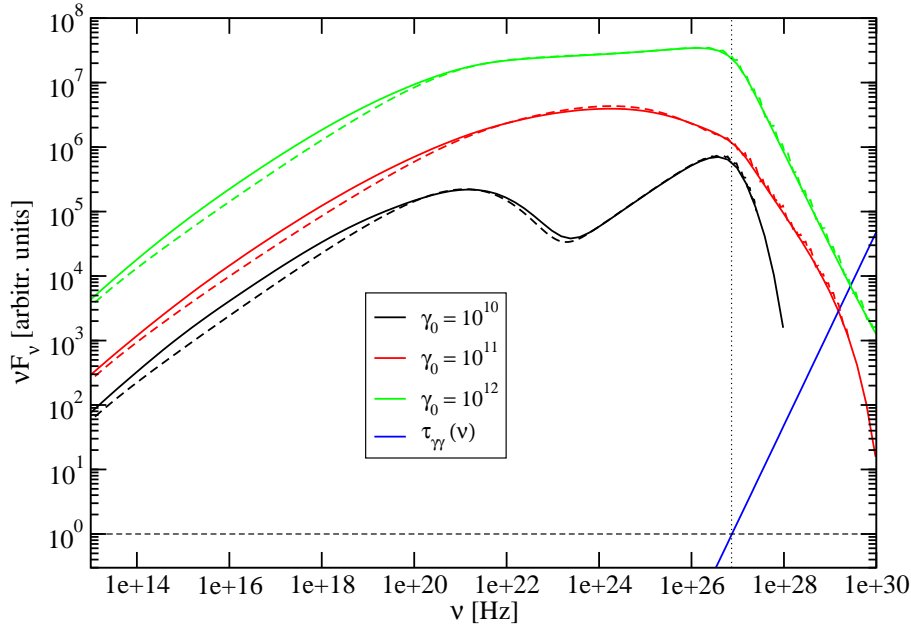}
\caption{\label{cascade_comparison}Comparison of the cascade
emission from Monte-Carlo simulations (solid curves) and our
semi-analytical description (dashed curves) in the case of
the injection of mono-energetic electrons with energies given
by the labels, in a $B = 10$~G magnetic field. The blue curve
shows the $\gamma\gamma$ opacity as a function of $\gamma$-ray
photon frequency. }
\end{figure}

Once the equilibrium pair distribution $N_e (\gamma)$ is known, it can be
used in Eq. \ref{Ndotsy} to evaluate the synchrotron emissivity and hence,
using Eq. \ref{Ndotescape} the observable photon spectrum. 

Figure \ref{cascade_comparison} illustrates the Monte-Carlo generated synchrotron 
+ cascade spectra (solid lines) for the injection of monoenergetic electrons 
with energies $\gamma_0 = 10^{10}$, $10^{11}$ and $10^{12}$, respectively. The
magnetic field is $B = 10$~G, and the $\gamma\gamma$ optical depth as a function
of photon energy is shown by the blue solid curve. The dashed lines illustrate
the results of the analytic approximations developed here. The agreement
between Monte-Carlo simulations and the semi-analytic approximation is
excellent, especially throughout the $\gamma$-ray regime. At lower energies,
our approximation remains accurate to within a factor of $\lesssim 2$.

\subsection{\label{protons}Proton Energy Losses and Equilibrium Particle Distribution}

In addition to the $\pi^0$ and electron/proton-synchrotron cascades,
proton synchrotron emission is the other main contributor to the high-energy
emission in hadronic models \citep{mp00,aharonian00}. In our model, we use an
asymptotic approximation analogous to Eq. (\ref{Ndotsy}) for the proton
synchrotron emissivity. 

An equilibrium proton distribution is evaluated assuming the injection
of a power-law distribution of protons, $Q_p (\gamma_p) = Q_{0, p}
\, \gamma_p^{-q_p} \, H (\gamma_p; \gamma_{1,p}, \gamma_{2,p})$. As
in the leptonic model described in \S \ref{lepton}, a self-consistent
equilibrium between this injection, particle escape, and cooling is
evaluated. Unlike the case of electrons, radiative cooling time scales
for protons can be longer than the typical dynamical time scale of the
expansion of the emission region. We therefore account for adiabatic
energy losses as well as radiative energy losses for the protons.

Adiabatic losses are evaluated through $\dot\gamma_p/\gamma_p = - \dot V
/ V$, where $V$ is the volume of the emission region. Assuming a conical
jet with opening angle $\theta_j \sim 1/\Gamma$, the adiabatic cooling
rate is $\dot\gamma_p = - 3 \, c \, \theta_j \, \gamma_p / R$. The
proton synchrotron cooling rate is 

\begin{equation}
\dot\gamma_{p, sy} = - {c \, \sigma_T \, B^2 \over 6 \pi \, m_e c^2}
\left( {m_e \over m_p} \right)^3 \, \gamma_p^2
\label{gdotpsy}
\end{equation}
and the photo-pion energy losses are approximated as \citep{aharonian00}

\begin{equation}
\dot\gamma_{p, p\gamma} = - c \langle \sigma_{p\gamma} f \rangle \,
n_{\rm ph} (\epsilon^{\ast}) \, \epsilon^{\ast} \; \gamma_p
\label{gdotpgamma}
\end{equation}
where $\langle \sigma_{p\gamma} f \rangle \approx 10^{-28}$~cm$^2$ is 
the elasticity-weighted $p\gamma$ interaction cross section and
$\epsilon^{\ast} = 5.9 \cdot 10^{-8} \, E_{19}^{-1}$ is the energy 
of target photons interacting with protons of energy $E$ at the 
$\Delta$ resonance, which, for plausible blazar parameters, is 
usually the most relevant contribution to the p$\gamma$ cross 
section \citep{muecke00b}.

\begin{figure}[ht]
\vskip 1cm
\centering
\includegraphics[width=12cm]{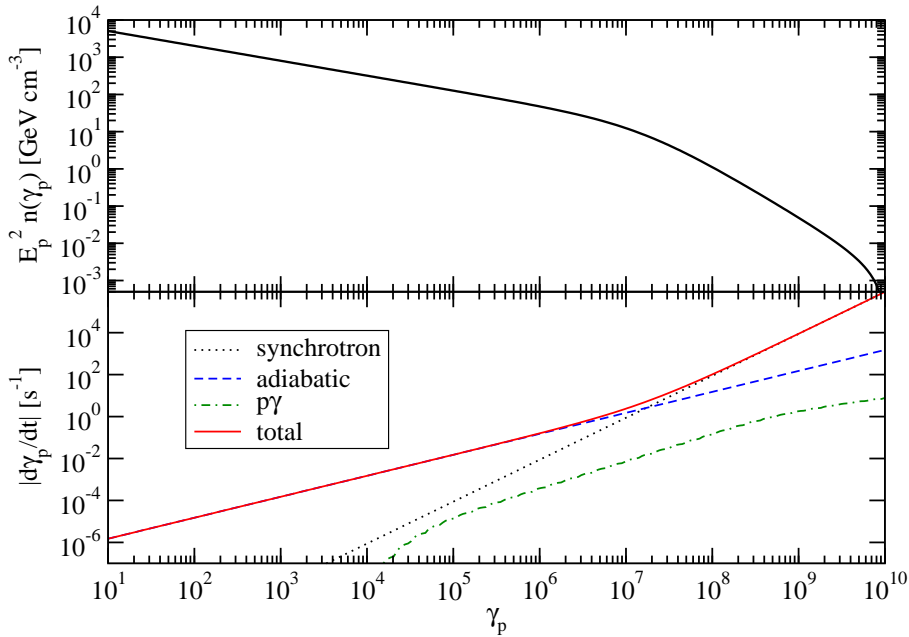}
\caption{\label{ploss}Proton cooling rates (bottom panel) and the 
equilibrium proton distribution (top panel) evaluated according to
Eq. (\ref{Np}) for typical synchrotron-proton-blazar model parameters
(see text for details). }
\end{figure}

The equilibrium proton distribution can then be found by integrating
the analog of the isotropic Fokker-Planck equation (\ref{fp}), neglecting
particle escape (which only affects the lowest-energy protons which
do not substantially contribute to the radiative output):

\begin{equation}
N_p (\gamma_p) = {-1 \over \dot\gamma_p} \int\limits_{\gamma}^{\infty}
d\tilde\gamma_p \, Q_p (\tilde\gamma_p)
\label{Np}
\end{equation}
In conical jet geometries, the energy loss rates of relativistic protons
can conceivably be dominated by adiabatic losses \citep[see, e.g.,][]{sikora11}.
Because of their linear dependence on $\gamma_p$, adiabatic losses will
not produce a break in the equilibrium proton spectrum. Eq. (\ref{Np})
predicts a break in the proton spectrum only where radiative (proton
synchroton or photo-pion production) losses become dominant. Figure 
\ref{ploss} illustrates this effect. The bottom panel shows the proton
energy loss rates due to adiabatic, synchrotron and photo-pion losses,
along with the total cooling rate. The target photon field is dominated
by the synchrotron emission from an electron population injected between
$\gamma_{e,1} = 10^3$ and $\gamma_{e,2} = 10^5$ with $q_e = 2.6$, in a magnetic
field of $B = 30$~G. Protons were injected with a power-law distribution
of index $q_p = 2.4$, between $E_{p,1} = 1$~GeV and $E_{p,2} = 10^{10}$~GeV.
The emission region has a radius of $R = 10^{16}$~cm and expands at an
angle of $\theta_{\rm j} = 1/\Gamma$ for $\Gamma = 20$. For these parameters,
radiative losses (dominated by proton synchrotron) begin to dominate over
adiabatic losses at $\gamma_{p, b} \sim 2 \times 10^7$. While for all proton 
Lorentz factors, the cooling time scale is shorter than the escape time scale,
a break in the equilibrium proton distribution (from index 2.4 to 3.4) 
occurs only at $\gamma_{p,b}$. 

The leptonic component of our hadronic jet model is being handled with 
the same procedure as described in \S \ref{lepton} for our leptonic model. 
The full leptonic radiative output (synchrotron + SSC) is used as target 
photon field for photo-pion production of the hadronic component and for 
evaluating the $\gamma\gamma$ opacity of the emission region.

\subsection{\label{limits}Limitations}

The \cite{ka08} templates we are using to evaluate the production spectra
of the final decay products, neglect the synchrotron (and Compton) emission
from intermediate decay products, i.e., muons and pions. It has been shown
\citep{muecke03} that under certain conditions, these emissions can make a
non-negligible contribution to the high-energy spectra in the SPB model. 
Neglecting these contributions is acceptable if one of two conditions
is fulfilled: (1) The synchrotron cooling time scale of relativistic
muons and pions are much longer than their decay time scales; or 
(2) proton synchrotron losses strongly dominate over photo-pion loses. 

For condition (1), we need to compare the decay time scales of pions (muons)
with Lorentz factors $\gamma_{\pi}$ ($\gamma_{\mu}$) in the blob rest frame, 

\begin{eqnarray}
t_d^{\pi} &=& 2.6 \times 10^{-8} \, \gamma_{\pi} \; {\rm s} \cr
t_d^{\mu} &=& 2.2 \times 10^{-6} \, \gamma_{\mu} \; {\rm s} \cr
\label{t_decay}
\end{eqnarray}

to the pion (muon) synchrotron cooling time scale analogous to 
Eq. (\ref{gdotpsy}). This yields the conditions

\begin{equation}
B \, \gamma_p \ll \cases{7.8 \times 10^{11} \; {\rm G} & for pions \cr
                         5.6 \times 10^{10} \; {\rm G} & for muons \cr}
\label{Bgammap}
\end{equation}
The Lorentz factors of secondary pions and muons are of the order of the
Lorentz factors of the primary protons. Therefore, Eq. (\ref{Bgammap})
constitutes a restriction to the highest proton Lorentz factors allowed
for a given magnetic field so that pion and muon synchrotron emission can
be neglected. Obviously, the condition is more restrictive for muons than
for pions. Our code automatically prints out a warning if the condition 
(\ref{Bgammap}) is not fulfilled for either muons or pions. 

To check for condition (2), i.e., proton synchrotron losses being dominant
over photo-pion losses, our code evaluates the proton synchrotron and  
photo-pion energy loss rates automatically (as plotted, for example, in
Figure \ref{ploss}). We only trust our results if either condition (1) for
muons, or condition (2) is fulfilled. Specifically, this means that our
semi-analytical hadronic model is not applicable to situations with both
very high magnetic fields and high radiation energy densities, in which
photo-pion production may dominate over proton synchrotron emission, and
the synchrotron cooling time scales of muons (and pions) may be shorter
than their decay time scale. 

In summary, our steady-state hadronic emission model is limited by its
neglect of pion and muon synchrotron radiation, implying restrictions
to the maximum possible field strength as a function of maximum proton
energy injected. Protons may lose their energy radiatively either via 
the proton synchrotron channel, or photomeson production. If the latter 
is the case, Bethe-Heitler pair production is considered subdominant, 
and is therefore neglected. The high energy photons and secondary 
particles from photomeson production may initiate cascades where the 
injected power is restributed by means of synchrotron radiation. We 
restrict ourselves to linear cascades. Inverse Compton losses are neglected,
owing to the large magnetic-field strengths in the emission region considered
here. 

External radiation fields as target field for particle and photon 
interactions and cascading are also not considered here. 
We point out that in the case of FSRQs, this simplification may not 
always be justified \citep[e.g.,][]{ad03}. We have therefore carefully 
verified whether the external radiation fields in the 6 FSRQs modelled in 
this paper (see \S \ref{modeling}) may make a substantial contribution to 
the target photon fields for pion production. For most of them, the 
co-moving synchrotron photon energy density in our hadronic model is 
at least two orders of magnitude larger than the co-moving-frame energy 
density of the external radiation field used for the leptonic models
(which were based on the observed BLR luminosity and a typical size 
of the BLR). Direct accretion disk emission is unlikely to play a
non-negligible role due to the unfavorable angle of incidence, which
requires very high proton Lorentz factors for photo-pion production,
as would the interaction with an infrared radiation field from a dust
torus. However, for 3C454.3 and PKS~0528+134, we find that the 
BLR radiation field (Doppler-boosted into the co-moving frame) 
may be of the order of or even larger than the synchrotron photon energy 
density, depending on the (unknown) radius of the BLR and the location
of the $\gamma$-ray emitting region with respect to the BLR. We therefore 
caution that our neglect of external radiation fields may not be valid 
in those two cases.

\section{\label{modeling}Comparative Modeling of {\it Fermi}-LAT 
detected blazars}

In this section, we will apply our leptonic and hadronic models described
in the previous sections, to a set of $\gamma$-ray blazars detected by
{\it Fermi}-LAT. \cite{abdo10} presented a comprehensive compilation of
simultaneous or contemporaneous broadband SEDs of a large sample of
{\it Fermi}-LAT detected blazars. Out of their list, we selected a
subset of blazars which fulfilled the following criteria: (1) good
simultaneous broadband SED coverage, including at least optical/IR, 
X-ray, and $\gamma$-ray data; (2) known redshift; (3) information 
about short-term variability; (4) apparent superluminal motion; 
(5) in the case of FSRQs: information about the accretion-disk and 
BLR luminosity. Condition (1) is necessary to allow for meaningful
SED fitting. Condition (2) is required to determine the overall luminosity
requirements of models. Condition (3) allows us to derive constraints on
the size of the emission region from the observed variability time scale
$t_{\rm var}$ via $R \le D \, c \, t_{\rm var} / (1 + z)$; (4) superluminal
motion with a speed $\beta_{\perp, \rm app}$ allows us to place a lower
limit on the bulk Lorentz factor of $\Gamma \ge \sqrt{ \beta_{\perp, \rm app}^2
+ 1}$; and (5) knowledge of the accretion-disk and BLR luminosities further
constrains the modeling of FSRQs in which these sources of external radiation
fields are generally believed to be important.

\begin{table*}
\centering
\small{
\begin{tabular}{ccccccc}
\hline
\hline
Object Name  & Type & z & $\beta_{\perp, \rm app}$ & $t_{\rm var}$   & $L_{\rm disk}$ [erg~s$^{-1}$] & $L_{\rm BLR}$ [erg~s$^{-1}$] \\
\hline
3C~273       & FSRQ & 0.158 & 13 ($^1$)           & 1 d ($^2$)      & $1.3 \times 10^{47}$ ($^3$)   & $9.1 \times 10^{45}$ ($^4$)  \\
3C~279       & FSRQ & 0.536 & 20.6 ($^5$)         & 2 d ($^6$)      & $2.0 \times 10^{45}$ ($^7$)   & $2.0 \times 10^{44}$ ($^8$)  \\
3C~454.3     & FSRQ & 0.859 & 14.9 ($^5$)         & 1 hr ($^9$)     & $1.7 \times 10^{47}$ ($^{10}$)& $2.5 \times 10^{45}$ ($^{11}$) \\
PKS~1510-089 & FSRQ & 0.36  & 20.3 ($^5$)         & 1 d ($^{12}$)   & $1.0 \times 10^{46}$ ($^{13}$)& --- \\
PKS~0420-01  & FSRQ & 0.916 & 7.5 ($^5$)          & 4 d ($^{14}$)   & $1.5 \times 10^{46}$ ($^{11}$)& $4.3 \times 10^{44}$ ($^{11}$) \\
PKS~0528+134 & FSRQ & 2.06  & 30 ($^{15}$)        & 1 d ($^{16}$)   & $1.7 \times 10^{47}$ ($^{16}$)& $1.8 \times 10^{46}$ ($^{16}$) \\
\hline
BL Lacertae  & LBL  & 0.069 & 10.7 ($^5$)         & 1.5 hr ($^{17}$)& $6.0 \times 10^{44}$ ($^{18}$)& $6.3 \times 10^{42}$ ($^{11}$) \\
AO~0235+164  & LBL  & 0.94  & 2 ($^5$)            & 1.5 d ($^{19}$) & $3.4 \times 10^{44}$ ($^{11}$)& $3.7 \times 10^{43}$ ($^{11}$) \\
S5~0716+714  & LBL  & 0.31  & 10.2 ($^5$)         & 15 min ($^{20}$)& ---                           & --- \\
OJ~287       & LBL  & 0.306 & 15.3 ($^5$)         & 2 hr ($^{21}$)  & $1.1 \times 10^{45}$ ($^{11}$)& $2.7 \times 10^{43}$ ($^{11}$) \\
\hline
W~Comae      & IBL  & 0.102 & 2 ($^{22}$)         & 5 hr ($^{23}$)  & $7.2 \times 10^{44}$ ($^{11}$)& $5.0 \times 10^{41}$ ($^{11}$) \\
3C~66A       & IBL  & 0.44 ?& 27 ($^{24}$)        & 6 hr ($^{25}$)  & ---                           & --- \\
\hline
\end{tabular}
\caption[]{\label{constraints}Observational data used to constrain our models. Superscripts
in paranetheses refer to the following references: 
1 = \cite{lister09}; 2 = \cite{courv88}; 3 = \cite{vf09}; 4 = \cite{peterson04}; 
5 = \cite{hovatta09}; 6 = \cite{boettcher07}; 7 = \cite{hartman01}; 8 = \cite{pian05}; 
9 = \cite{raiteri08b}; 10 = \cite{raiteri08c}; 11 = \cite{xie08}; 12 = \cite{marscher10};
13 = \cite{pucella08}; 14 = \cite{darcangelo07}; 15 = \cite{jorstad05}; 16 = \cite{palma11}; 
17 = \cite{villata02}; 18 = \cite{raiteri09}; 19 = \cite{raiteri08a}; 20 = \cite{sasada08}; 
21 = \cite{fan09}; 22 = \cite{massaro01}; 23 = \cite{tagliaferri00}; 24 = \cite{jorstad01};
25 = \cite{takalo96}
}}
\end{table*}

We found that the following objects fulfilled our criteria: The FSRQs 
3C~273, 3C~279, 3C~454.3, PKS~1510-089, PKS~0420-01, and PKS~0528+134;
the LBLs BL~Lacertae, AO~0235+164, S5~0716+714, and OJ~287; and the
IBLs W~Comae and 3C~66A. SED data for these objects are obtained from
\cite{abdo10}, and Table \ref{constraints} lists observational data
that are used to further constrain our models.

Our fitting procedure is a ``fit-by-eye'' method, starting with plausible 
values for the parameters that are not directly constrained by observations.
The unconstrained parameters are then adjusted to obtain an acceptable
representation of the SED. Simpler models with fewer parameters, such as
a single-zone, static SSC model, allow for a detailed $\chi^2$ minimization 
technique \citep[see, e.g.,][]{mank11}. However, given the substantial number
of adjustable parameters, such a fitting method is infeasible for the models
considered here. While we are confident that our best-fit parameters are
in the correct ball-park range for these objects, the lack of a rigorous
$\chi^2$ minimization strategy prevents us from determining errors on the
fit parameters. Therefore, while we will proceed to a global assessment
of parameter values among different classes of blazars, we will refrain
from any quantitative statements in this respect. 

In the course of the fitting, we strive to achieve acceptable fits with 
parameters close to equipartition between the dominant particle species 
(electrons in the leptonic model, protons in the hadronic model) and the 
magnetic field. This is motivated by two arguments: (1) If the relativistic
jets of AGN are powered by rotational energy from the central black hole
\citep{bz77}, the jets are expected to be initially Poynting-flux dominated,
and the energy carried in electromagnetic fields needs to be transferred to
relativistic particles in order to produce the observed high-energy emission.
This energy conversion is expected to cease as approximate equipartition
between the magnetic field and relativistic particles is reached, so that
the jets are not expected to become matter dominated in the central few
parsecs of the AGN, where the high-energy emission in blazars is believed
to be produced \citep{lyubarsky10}. (2) If magnetic pressure plays an
essential role in collimating AGN jets out to kpc scales, the particle
pressure can not largely dominate over the magnetic pressure in the
inner few parsecs of the AGN. For these reasons, we prefer model parameters
with magnetic fields dominating the pressure and energy density in the
emission region over particle-dominated scenarios. 

\begin{figure}
\bigskip
\bigskip
\begin{center}
\vbox{\hbox{\includegraphics[width=6.8cm]{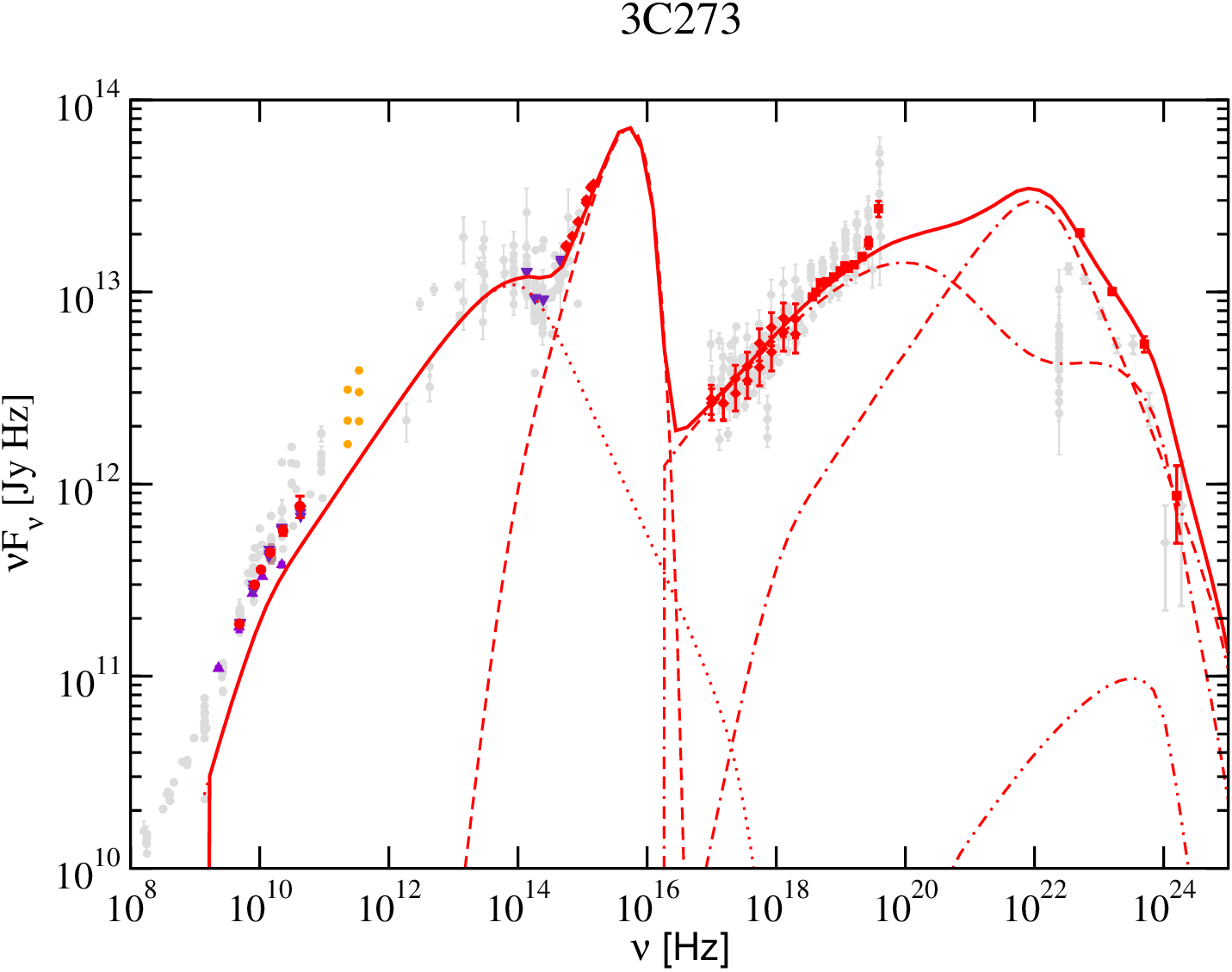}
            \hskip 1cm
            \includegraphics[width=6.8cm]{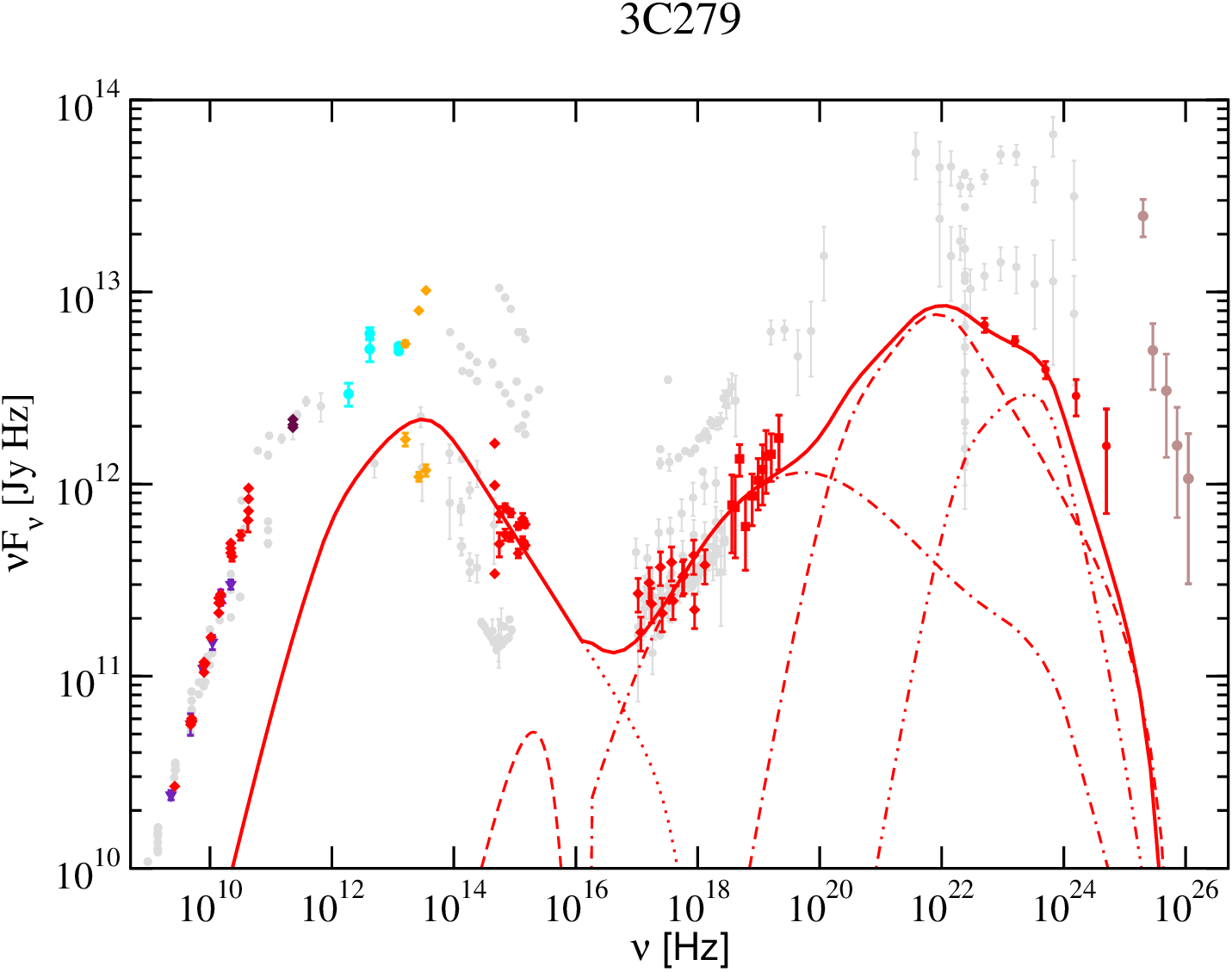}
            }
      \bigskip
      \hbox{\includegraphics[width=6.8cm]{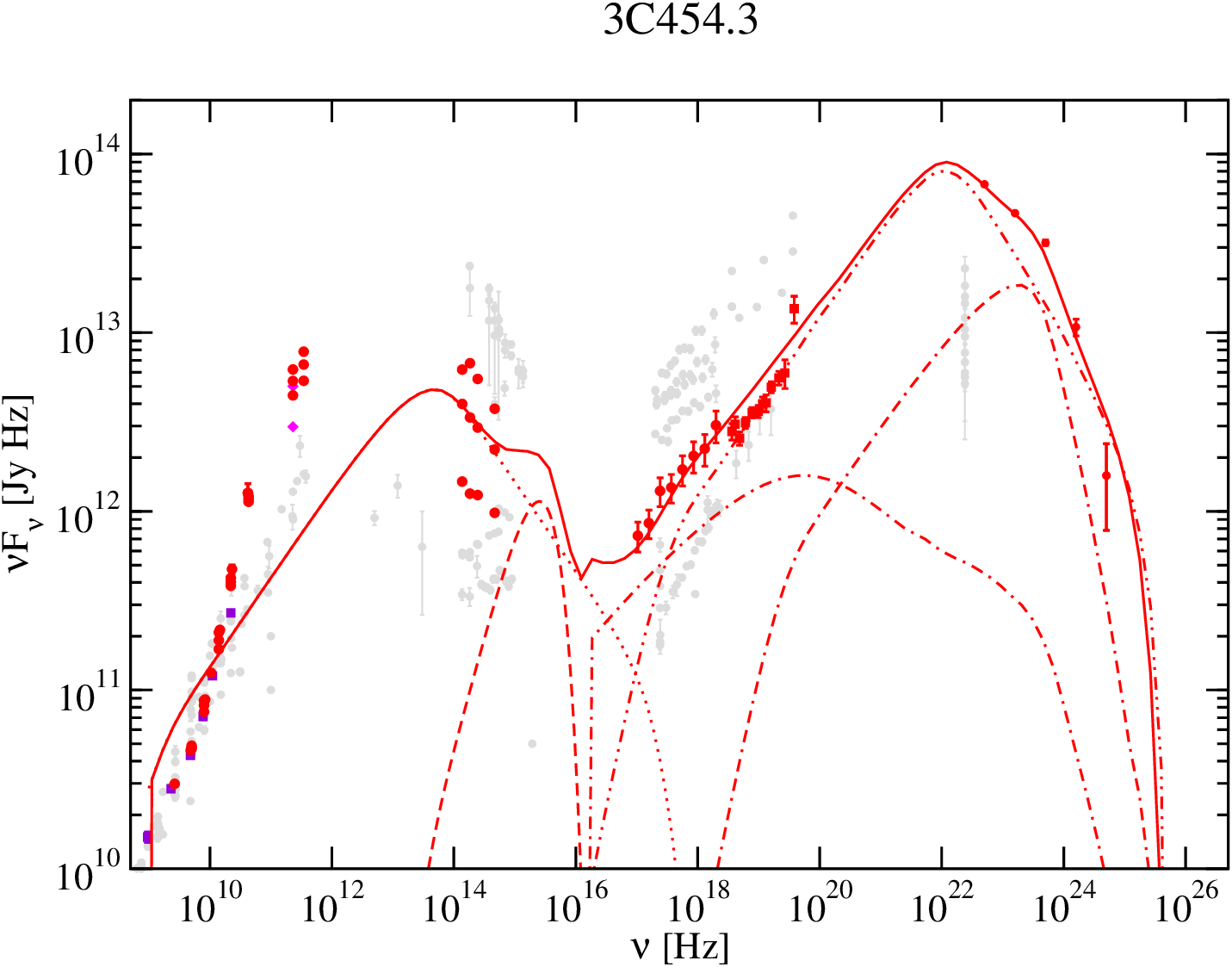}
            \hskip 1cm
            \includegraphics[width=6.8cm]{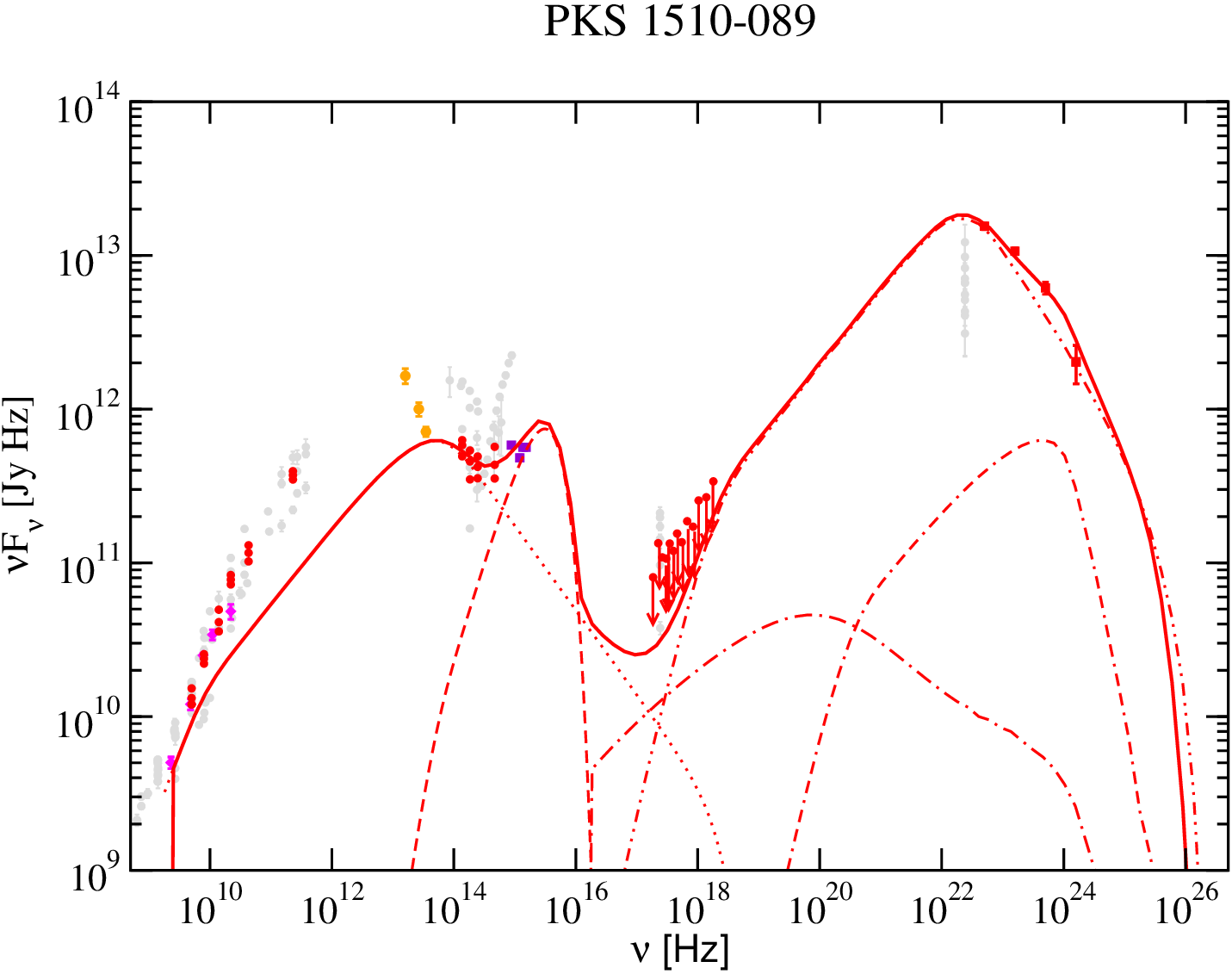}
            }
      \bigskip
      \hbox{\includegraphics[width=6.8cm]{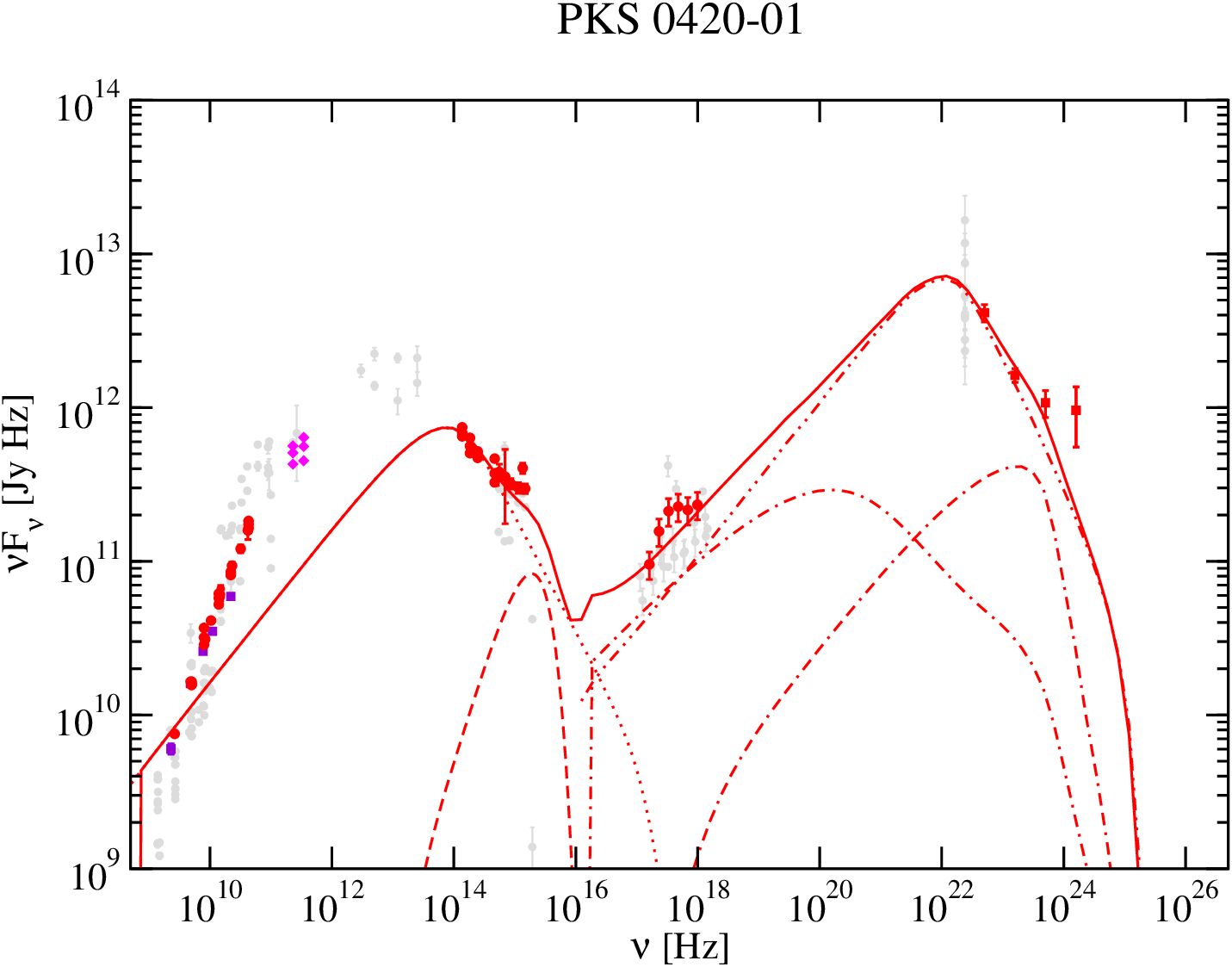}
            \hskip 1cm
            \includegraphics[width=6.8cm]{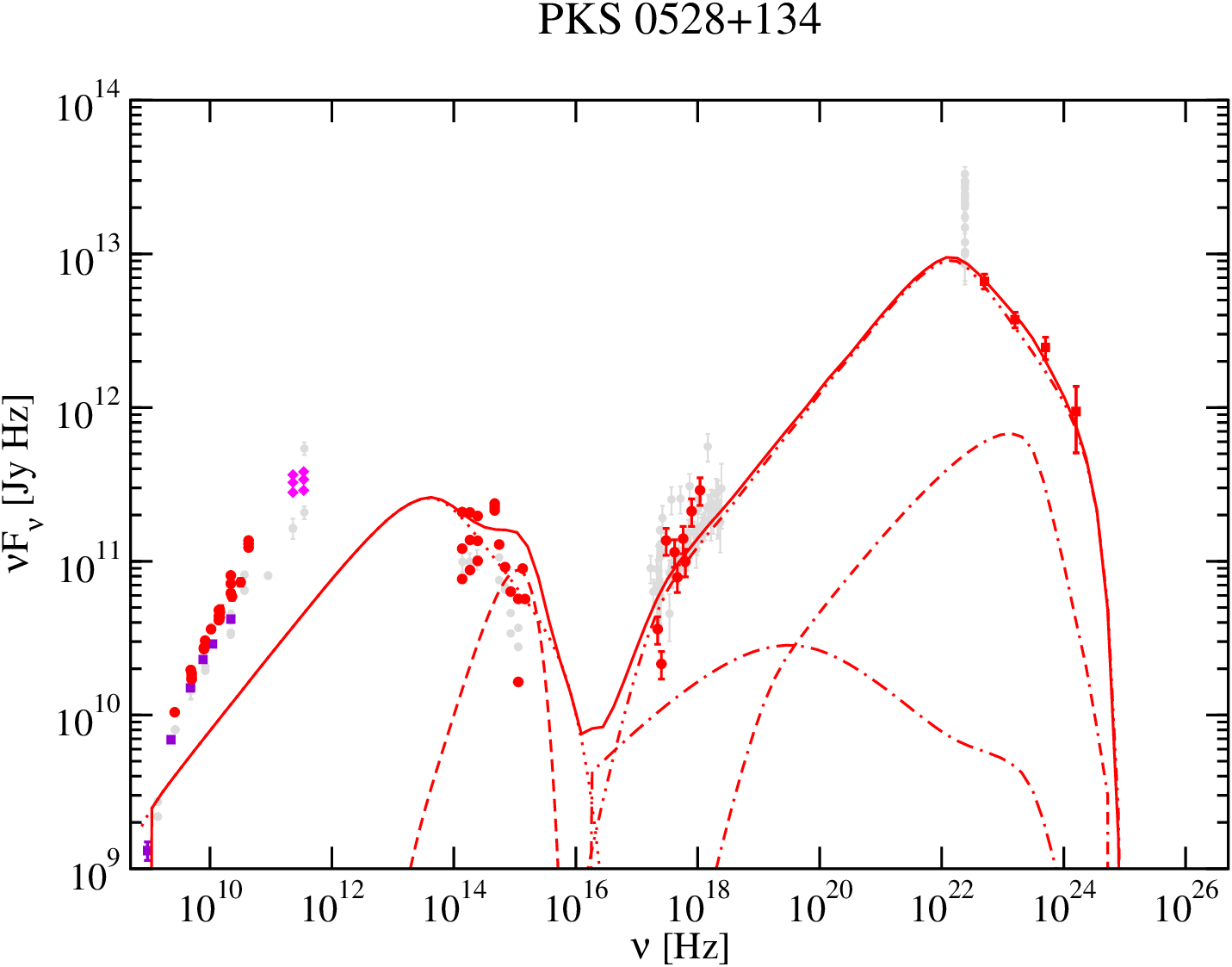}
            }
      }
\end{center}
\caption{\label{leptonic_FSRQ}Leptonic model fits to the 6 FSRQs in
our sample. See table \ref{fitpars} for parameters. Dotted = synchrotron;
dashed = accretion disk; dot-dashed = SSC; dot-dash-dashed = EC (disk);
dot-dot-dashed = EC (BLR). }
\end{figure}

For all objects, we produced one leptonic and one hadronic model 
fit to the contemporaneous SED. In the case of 3C~66A, the catalog
value of the redshift of $z = 0.444$ is highly questionable. The recent
analyses of \cite{prandini10} and \cite{3C66A} place the object at a likely
redshift of $z \sim 0.2$ -- 0.3, while \cite{furniss13} provide a lower 
limit on the redshift of $z \ge 0.3347$, based on UV absorption features 
by intervening intergalactic medium. In our modeling, we assume a fiducial
redshift of $z = 0.3$ for 3C~66A. While, of course, the parameter values 
slightly change when using a slightly higher redshift (say, $z = 0.35$ to 
be consistent with \cite{furniss13}), the overall conclusions from the 
modeling remain unaffected. All our model SEDs are corrected for 
$\gamma\gamma$ absorption by the extragalactic background light using 
the model of \cite{finke10}.

\begin{rotate}
\begin{table*}
\centering
\small{
\begin{tabular}{cccccccccccc}
\hline
\hline
Object       & $\gamma_{e,1}$  & $\gamma_{e,2}$  & $q_e$ & $B$ [G] & $D$ & $f_{\rm BLR}^a$      & $L_e^b$ & $L_p^b$ & $L_B^b$ & $\epsilon_{Be}$ & $t_{\rm var, min}^c$ \\
\hline
3C~273       & $1 \times 10^3$ & $5 \times 10^4$ & 3.5   & 2.0     & 13  & $5.4 \times 10^{-5}$ &  6.0    & 130     & 0.63    & 0.11            & 4.1 \\
3C~279       & $1 \times 10^3$ & $1 \times 10^5$ & 3.0   & 0.7     & 17  & $1.7 \times 10^{-2}$ &  5.8    & 29      & 5.4     & 0.94            & 29  \\
3C~454.3     & 800             & $5 \times 10^4$ & 3.0   & 2.1     & 15  & $8.0 \times 10^{-2}$ &  22     & 870     & 75      & 3.5             & 52  \\
PKS~1510-089 & $1 \times 10^3$ & $2 \times 10^5$ & 3.1   & 0.8     & 20  & $4.1 \times 10^{-3}$ &  2.1    & 41      & 2.1     & 1.0             & 9.4 \\
PKS~0420-01  & $1.4 \times 10^3$ & $5 \times 10^4$ & 3.4 & 2.5     & 8.0 & $1.0 \times 10^{-1}$ &  6.9    & 670     & 38      & 5.4             & 110 \\
PKS~0528+134 & 700             & $1 \times 10^4$ & 3.0   & 3.0     & 19  & $9.1 \times 10^{-3}$ &  9.4    & 600     & 110     & 11              & 45  \\
\hline
BL Lacertae  & $1.1 \times 10^3$ & $1 \times 10^5$ & 3.2 & 2.5     & 15  & $1.3 \times 10^{-1}$ &  0.44   & 4.6     & 0.41    & 0.94            & 1.8 \\
AO~0235+164  & 700            & $5 \times 10^4$ & 3.7   & 1.3     & 25  & $1.1$                &  12     & 87      & 33      & 2.8             & 21  \\
S5~0716+714  & $1.9 \times 10^3$ & $5.5 \times 10^4$ & 3.8 & 3.8   & 20  & --- $^d$             &  1.3    & 6.7     & 14      & 10              & 4.9 \\
OJ~287       & 800             & $5 \times 10^4$   & 3.8 & 3.5     & 15  & $1.5 \times 10^{-1}$ &  1.8    & 16      & 15      & 8.2             & 9.7 \\
\hline
W~Comae      & $1 \times 10^3$ & $8 \times 10^4$   & 2.4 & 1.5     & 30  & $2.8 \times 10^{-2}$ &  0.30   & 1.52    & 0.30    & 1.0             & 0.68\\
3C~66A       & $9 \times 10^3$ & $3 \times 10^5$   & 2.8 & 0.065   & 40  & --- $^d$             &  13     & 8.2     & 0.45    & 0.034           & 13  \\
\hline
\end{tabular}
\caption[]{\label{fitpars}Parameters for the leptonic SED fits shown in Figures \ref{leptonic_FSRQ} -- \ref{leptonic_IBL}. 
$^a$ The factor $f_{\rm BLR}$, determining the energy density of the re-processed disk radiation field, is defined as 
$f_{\rm BLR} \equiv \eta_{\rm BLR} / R_{\rm BLR}^2$; values are in pc$^{-2}$. $^b$ Powers are in units of 
$10^{44}$~erg~s$^{-1}$; $^c$ minimum allowed variability time scale in hours. $^d$ For S5~0716+714 and 3C~66A, no accretion 
disk luminosity has been measured/constrained; For S5~0716+714, an isotropic external radiation field with $u_{\rm ext} = 
5 \times 10^{-5}$~erg~cm$^{-3}$ and $T_{\rm BB} = 2 \times 10^4$~K has been used for the SED fit; for 3C~66A:
$u_{\rm ext} = 1.3 \times 10^{-8}$~erg~cm$^{-3}$ and $T_{\rm BB} = 10^3$~K. 
}}
\end{table*}
\end{rotate}

\subsection{\label{leptonic_fits}Leptonic Model Fits}

\begin{figure}
\bigskip
\bigskip
\begin{center}
\vbox{\hbox{\includegraphics[width=6.8cm]{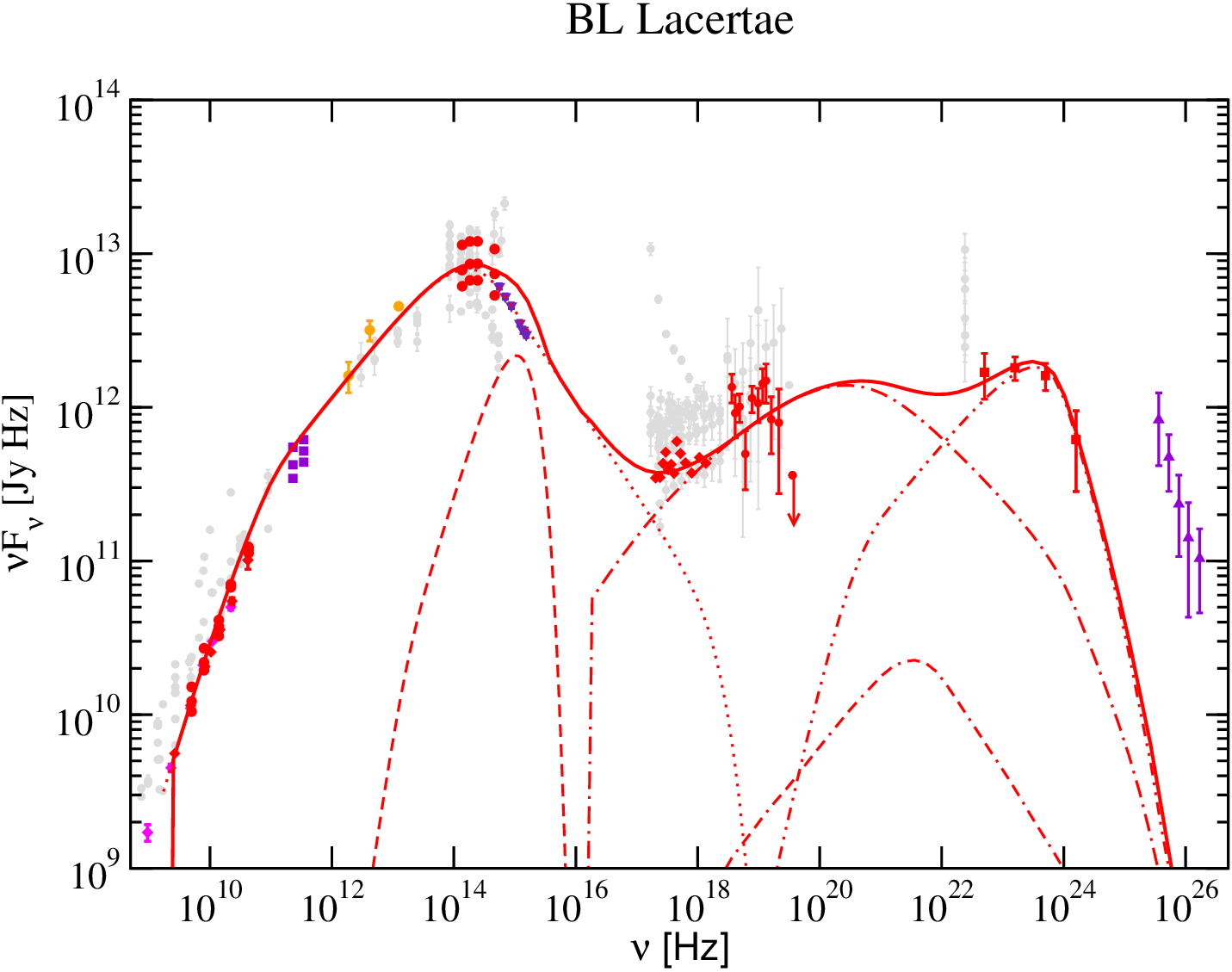}
            \hskip 1cm
            \includegraphics[width=6.8cm]{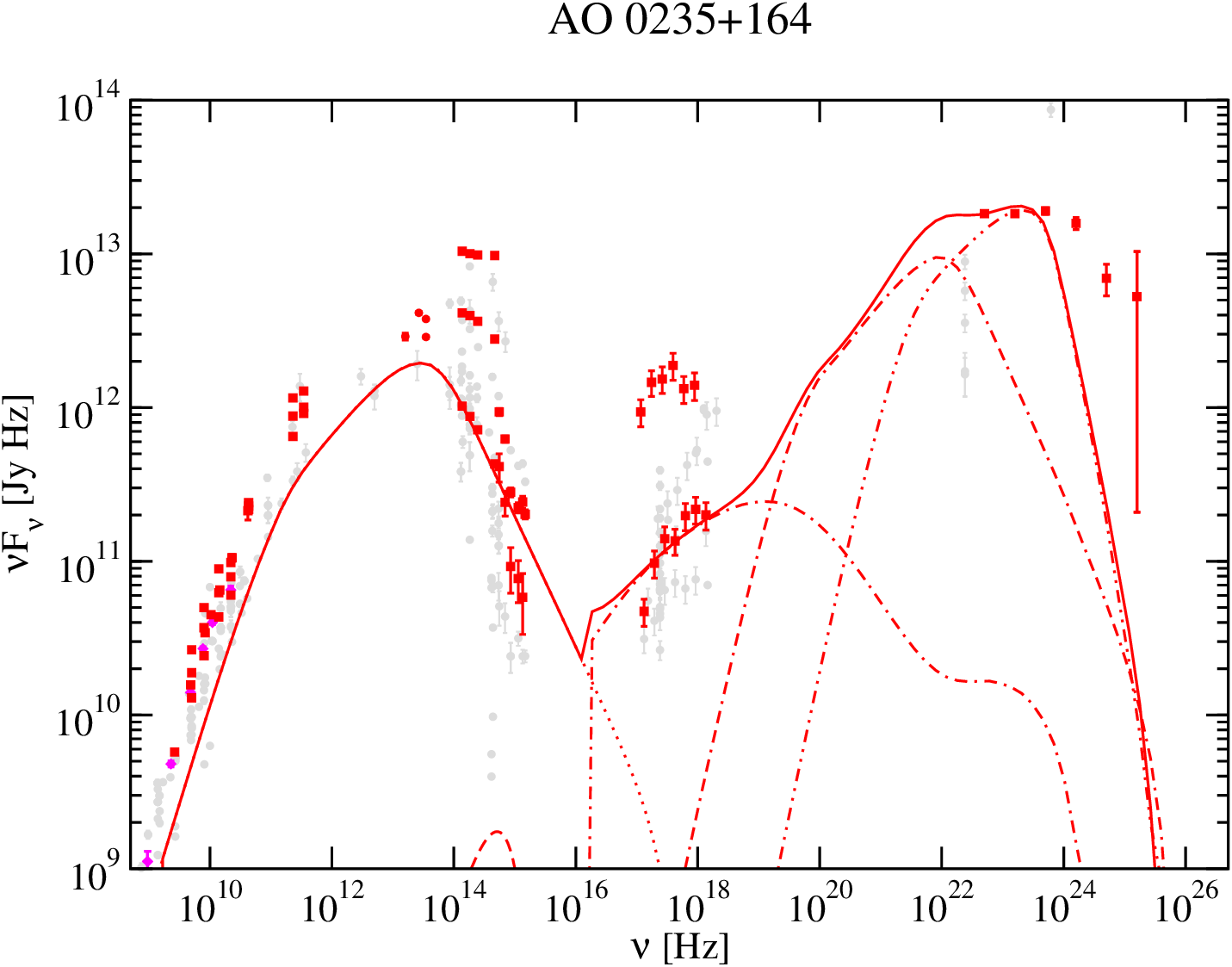}
            }
      \bigskip
      \hbox{\includegraphics[width=6.8cm]{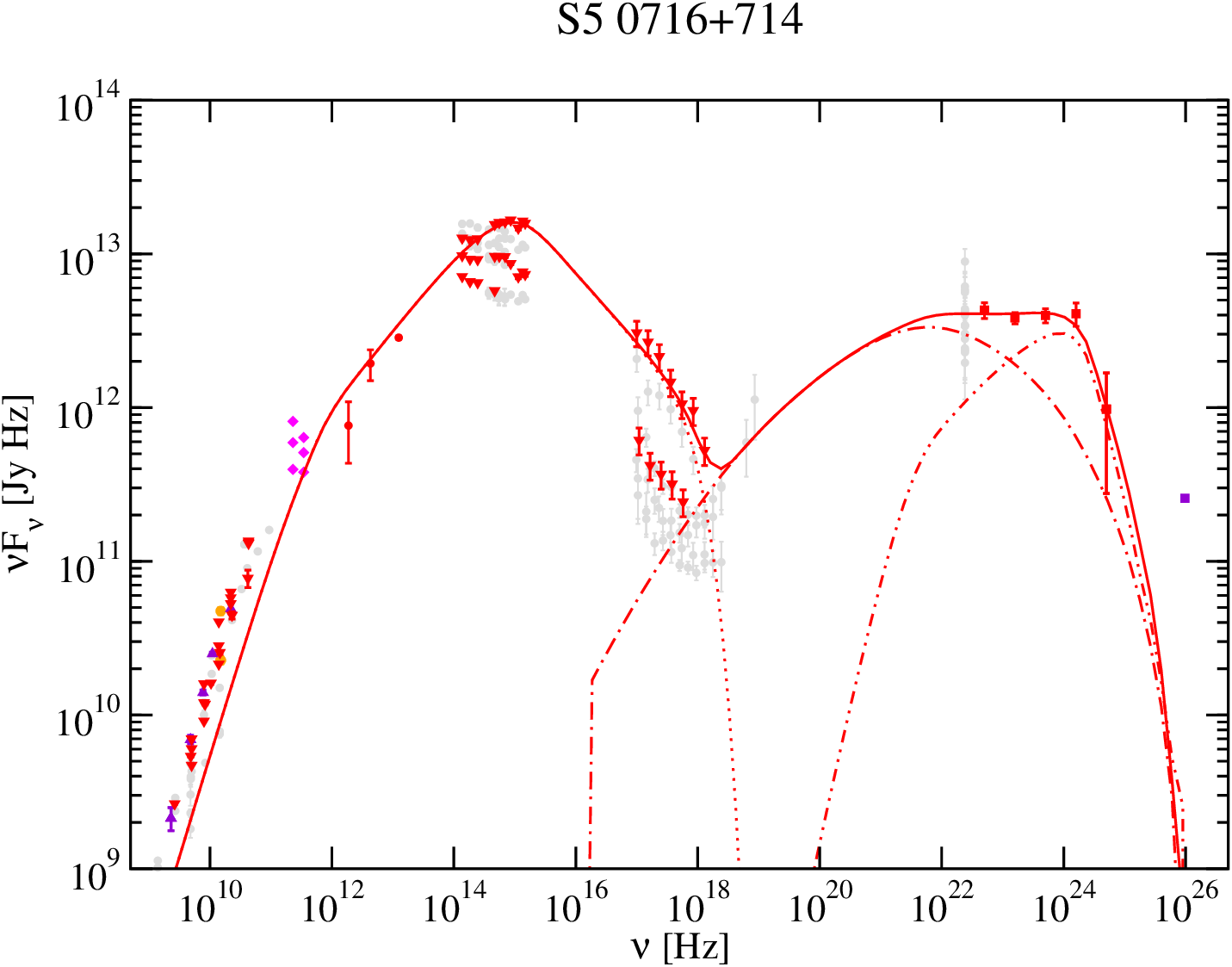}
            \hskip 1cm
            \includegraphics[width=6.8cm]{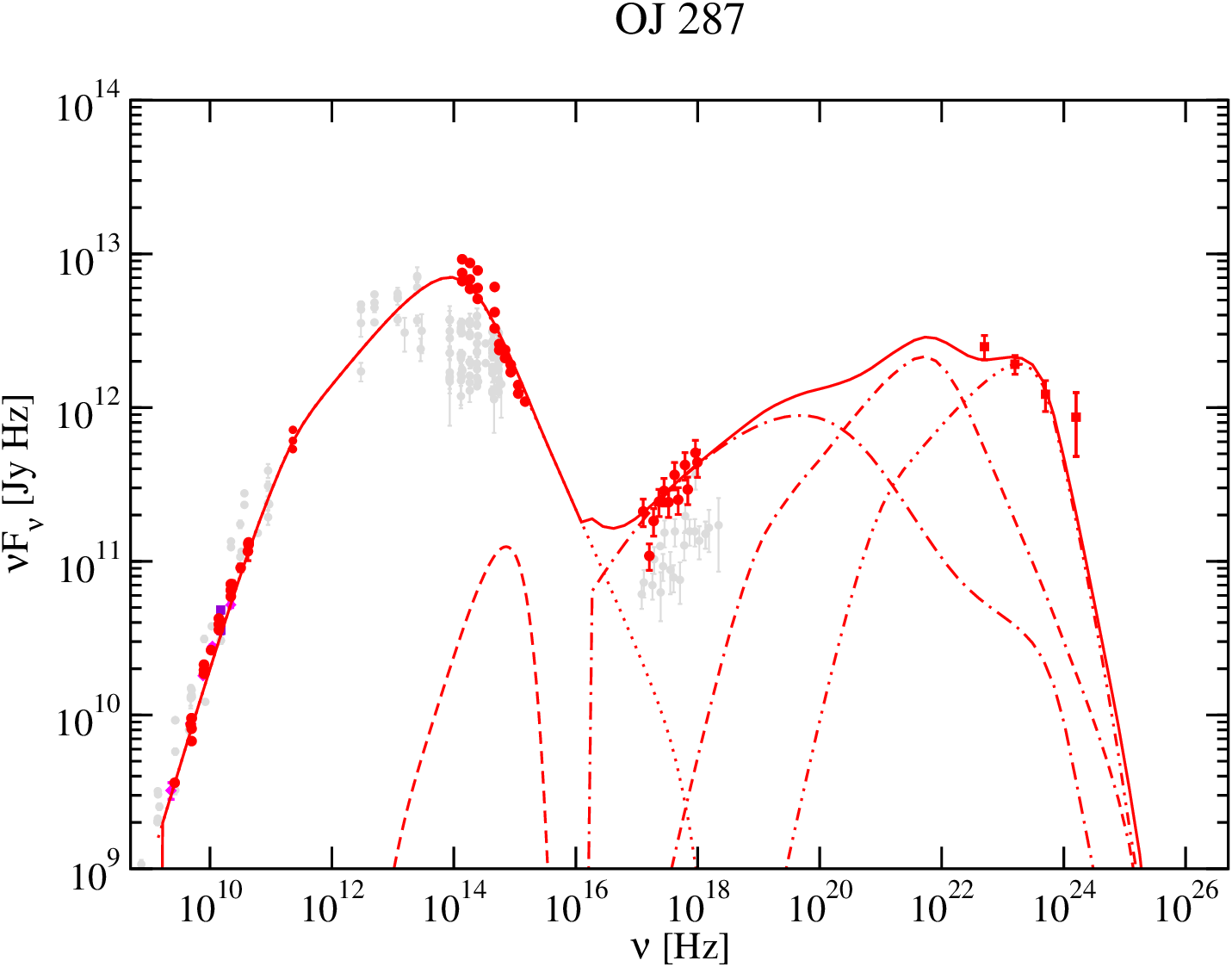}
            }
      }
\end{center}
\caption{\label{leptonic_LBL}Leptonic model fits to the 4 LBLs in
our sample. See table \ref{fitpars} for parameters. Dotted = synchrotron;
dashed = accretion disk; dot-dashed = SSC; dot-dash-dashed = EC (disk);
dot-dot-dashed = EC (BLR). }
\end{figure}

\begin{figure}
\bigskip
\bigskip
\begin{center}
\hbox{\includegraphics[width=6.8cm]{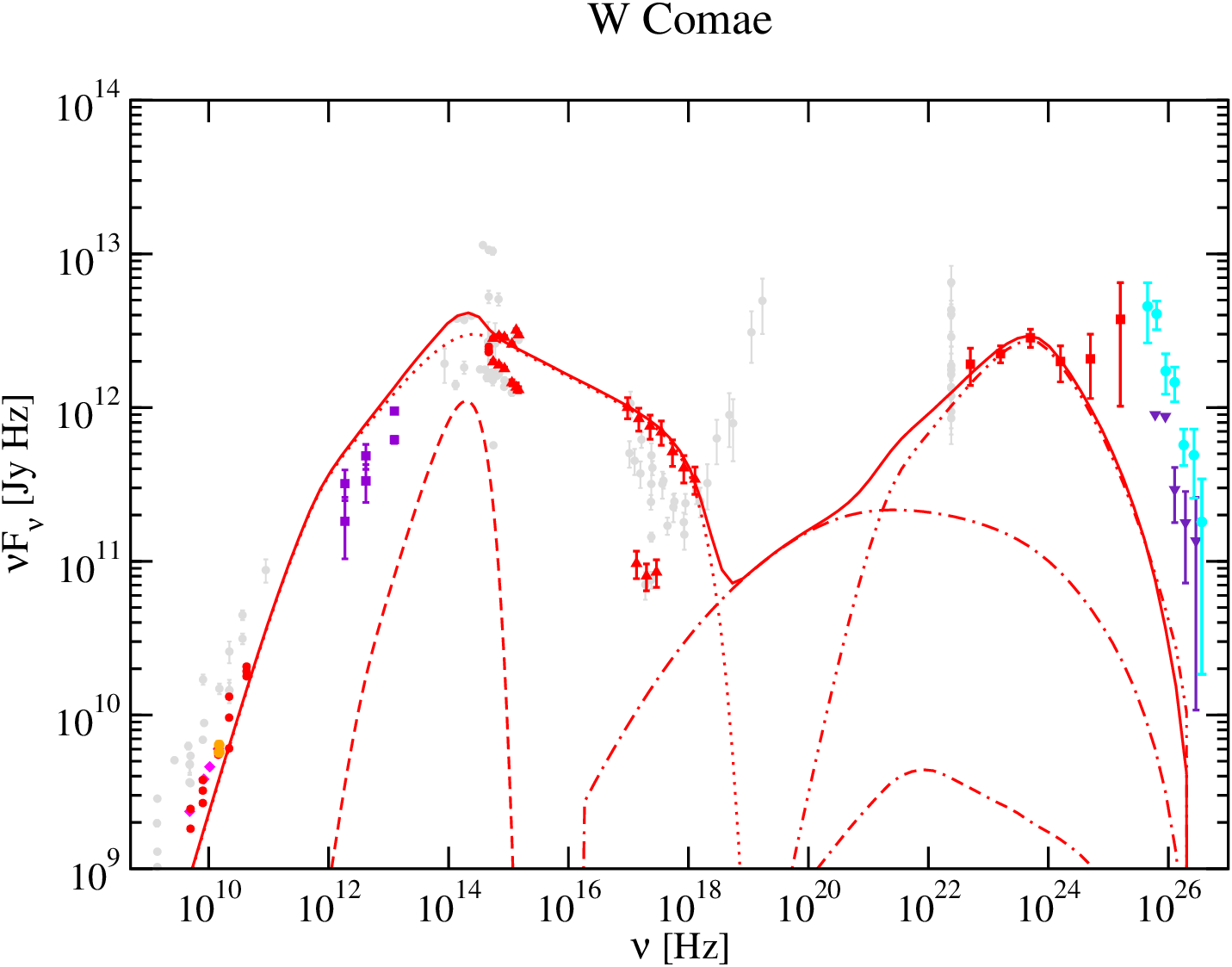}
      \hskip 1cm
      \includegraphics[width=6.8cm]{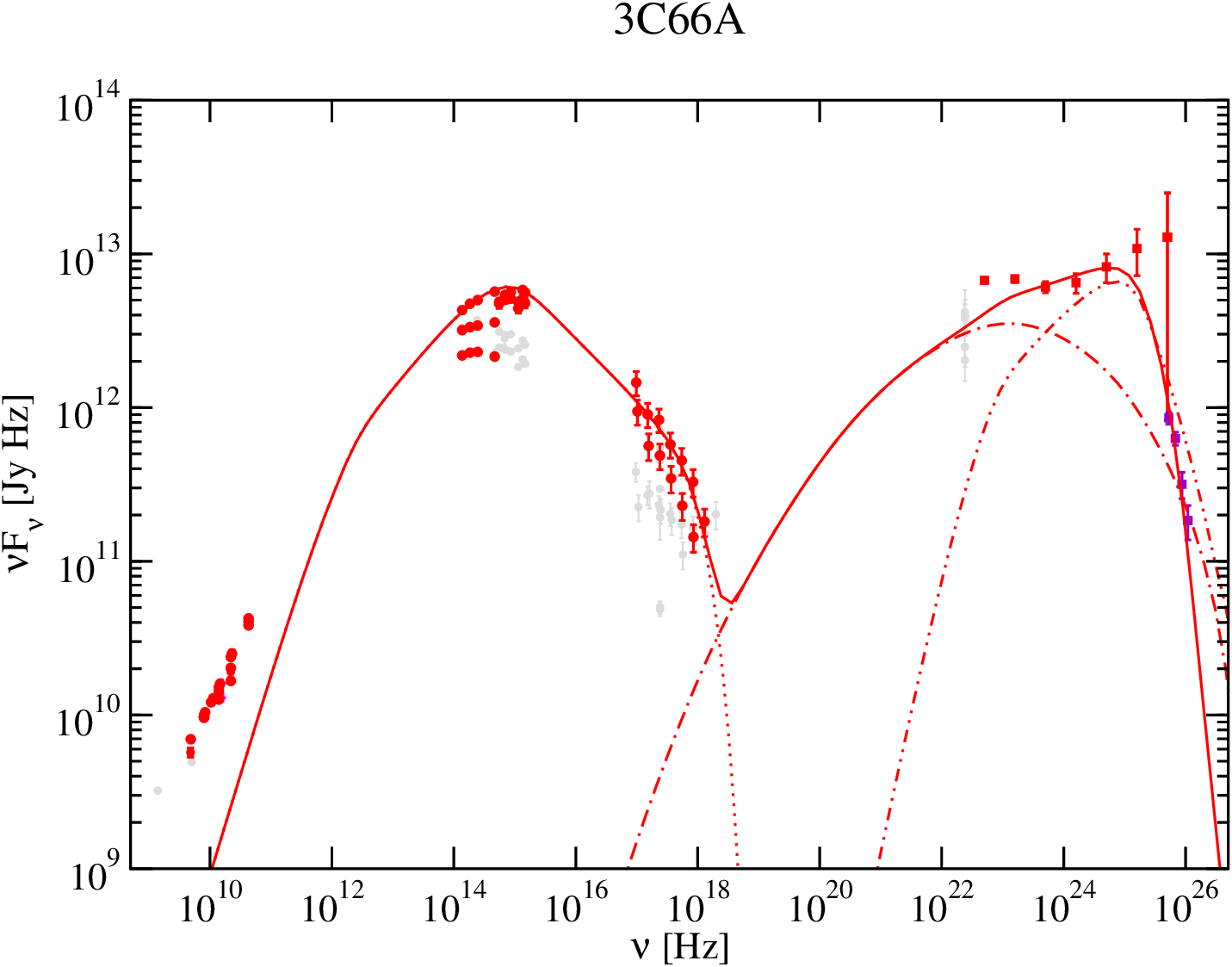}
      }
\end{center}
\caption{\label{leptonic_IBL}Leptonic model fits to the 2 IBLs in
our sample. See table \ref{fitpars} for parameters. Dotted = synchrotron;
dashed = accretion disk; dot-dashed = SSC; dot-dash-dashed = EC (disk);
dot-dot-dashed = EC (BLR). }
\end{figure}

Figures \ref{leptonic_FSRQ} -- \ref{leptonic_IBL} show the SEDs and our
leptonic model fits for the 12 blazars we selected for this study. The fit
parameters used and a few quantities derived from those parameters, are
listed in Table \ref{fitpars}. In general, satisfactory fits to most blazar 
SEDs with parameters close to (within an order of magnitude of) equipartition 
can be achieved with the leptonic model described above. 

In agreement with many previous studies, we find that FSRQs require a 
dominant contribution from external-Compton (on the direct accretion 
disk and the BLR emission) in order to provide acceptable SED fits 
\citep[e.g.,][]{sambruna97,ghisellini98,mukherjee99,hartman01}. In 
our model fits, we are able to produce the spectral breaks observed 
in the Fermi-LAT spectra of many blazars, with a superposition of 
two different $\gamma$-ray emission components, as previously proposed 
for the case of 3C454.3 by \cite{fd10}. Typical minimum variability
time scales from light-travel time arguments are of the order of
1 -- 2 days, consistent with the day-scale variability seen in most
Fermi-detected FSRQs.

As argued previously, e.g. by  \cite{madejski99} and \cite{bb00} for
the case of BL~Lacertae, also LBLs are represented with more plausible
parameters when including an EC component using a low-luminosity accretion
disk and BLR, compared to a pure SSC model. In the case of S5~0716+714,
where no observationally motivated estimates of the luminosity of the
accretion disk or the BLR could be found, we have assumed the existence 
of a dust torus with a very low infrared luminosity, consistent with the 
absence of any direct observational evidence for it. Our models allow for
variability on time scales of a few hours for the LBLs modelled here.
While the SEDs of BL Lacertae, S5~0716+714 and OJ~287 are well represented
by a leptonic SSC+EC model, our model has problems representing the 
very hard $\gamma$-ray spectrum of AO~0235+164 above a few GeV as our
model $\gamma$-ray spectra are truncated by Klein-Nishina effects. 

In blazars in which we find that a substantial contribution from EC on
direct accretion-disk emission is relevant, we find a best-fit distance
of the $\gamma$-ray emission region from the central black hole in the
range $z_0 \lesssim 0.1$~pc. An important contribution from EC on BLR
emission would imply a characteristic distance of $z \lesssim 1$~pc,
while a much larger distance would be consistent with situations in
which EC on infrared emission (from warm dust) dominates the $\gamma$-ray
production. As described in \S \ref{lepton}, the external radiation fields
from the BLR and/or warm dust are modelled as isotropic in the AGN rest 
frame so that for such fits, the distance to the black hole is not 
necessary input parameter for our model. 

While BL~Lac objects detected at $> 100$~GeV $\gamma$-rays by ground-based
Atmospheric Cherenkov Telescopes have often been found to be well represented
by pure SSC models, the careful analysis of the simultaneous SEDs of the
VERITAS-detected IBLs W~Comae and 3C~66A \citep{WComb,3C66A} revealed 
that also in these two cases, a pure SSC fit would require rather extreme
parameters with magnetic fields far below equipartition. This situation
could be remedied allowing for a contribution from external Compton to
the $\gamma$-ray emission. These findings are confirmed in this study.
However, the unusual, apparent upward curvature of the {\it Fermi}-LAT
spectrum of 3C~66A and W~Comae can not be satisfactorily represented with our model,
irrespective of the dominant target photon field for external Compton
scattering.

We find that the two IBLs in our sample require systematically higher
Doppler factors than the LBLs and FSRQs. Also, as expected, BL~Lac
objects are characterized by less powerful jets (i.e., lower $L_e$) 
than FSRQs, and the two IBLs require harder electron spectra than the 
LBLs and FSRQs. We do not find a systematic difference in magnetic-field 
values between BL~Lacs and FSRQs, neither are the characteristic electron
energies (represented by $\gamma_{e,1}$) systematically different between
the two classes of objects. If charge neutrality in blazar jets is provided
by cold protons (rather than positrons), our fits indicate that the kinetic
energy carried by the jets should be dominated by protons.

\begin{rotate}
\begin{table}
\centering
\tiny{
\begin{tabular}{cccccccccccccc}
\hline
\hline
Object       & $\gamma_{e,1}$ & $\gamma_{e,2}$ & $q_e$ & $B^a$ & $D$ & $\gamma_{\rm p, max}$ & $q_p$ & $L_e^b$ & $L_p^c$ & $\epsilon_{Be}^d$ & $\epsilon_{Bp}^d$    & $\epsilon_{ep}^d$    & $t_{\rm var}^e$ \\
\hline
3C~273       & 350            & $1.5 \times 10^4$ & 3.4 & 15 & 15  & $4.3 \times 10^8$       & 2.4   & 0.13    & 25      & 3300              & $1.7 \times 10^{-3}$ & $5.2 \times 10^{-7}$ & 11  \\
3C~279       & 100            & $2.0 \times 10^4$ & 3.0 & 100 & 15 & $6.4 \times 10^8$       & 2.2   & 0.19    & 3.5     & 1410              & $7.9 \times 10^{-3}$ & $5.6 \times 10^{-6}$ & 1.7 \\
3C~454.3     & 300            & $1.5 \times 10^4$ & 3.2 & 10 & 15  & $1.1 \times 10^9$       & 2.1   & 1.0     & 35      & $1.2 \times 10^4$ & 0.035                & $2.9 \times 10^{-6}$ & 138 \\
PKS~1510-089 & 150            & $1.5 \times 10^4$ & 3.2 & 10 & 20  & $1.1 \times 10^9$       & 1.7   & 0.57    & 2.5     & 24                & $5.4 \times 10^{-4}$ & $2.3 \times 10^{-5}$ & 1.9 \\
PKS~0420-01  & 75             & $1.0 \times 10^4$ & 3.2 & 100 & 10 & $4.3 \times 10^8$       & 1.3   & 0.52    & 0.42    & 2200              & 0.27                 & $1.2 \times 10^{-4}$ & 9.7 \\
PKS~0528+134 & 150            & $1.0 \times 10^4$ & 3.8 & 30 & 20  & $1.1 \times 10^9$       & 2.0   & 1.9     & 44      & 1020              & $4.4 \times 10^{-3}$ & $4.3 \times 10^{-6}$ & 17  \\
\hline
BL Lacertae  & 700            & $1.5 \times 10^4$ & 3.5 & 10 & 15  & $1.9 \times 10^9$       & 2.4   & 0.087   & 9.8     & 39                & $3.4 \times 10^{-5}$ & $8.9 \times 10^{-7}$ & 1.3 \\
0235+164     & 200            & 750               & 3.0 & 15 & 25  & $4.3 \times 10^9$       & 1.9   & 1.5     & 10      & 130               & $1.9 \times 10^{-3}$ & $1.5 \times 10^{-5}$ & 4.3 \\
S5~0716+714  & 900            & $3.0 \times 10^4$ & 2.9 & 20 & 15  & $2.7 \times 10^9$       & 2.0   & 0.089   & 0.14    & $1.4 \times 10^4$ & 0.85                 & $6.2 \times 10^{-5}$ & 15  \\
OJ~287       & 350            & $4.0 \times 10^4$ & 4.1 & 20 & 15  & $1.0 \times 10^9$       & 1.6   & 0.53    & 8.3     & 66                & 0.042                & $6.3 \times 10^{-4}$ & 2.6 \\
\hline
W~Comae      & 800            & $2.1 \times 10^4$ & 2.6 & 30 & 15  & $1.9 \times 10^9$       & 2.0   & 0.014   & 0.021   & 560               & 0.037                & $6.6 \times 10^{-5}$ & 0.68 \\
3C~66A       & 750            & $1.3 \times 10^4$ & 2.8 & 10 & 30  & $1.2 \times 10^9$       & 2.0   & 0.32    & 1.2     & 24                & $6.5 \times 10^{-4}$ & $2.7 \times 10^{-5}$ & 0.57 \\
\hline
\end{tabular}
\caption[]{\label{hfitpars}Parameters for the hadronic SED fits shown in Figures \ref{hadronic_FSRQ} -- \ref{hadronic_IBL}.  
$^a$ Magnetic field in units of Gauss. $^b$ Kinetic luminosity in relativistic electrons in units of $10^{44}$~erg~s$^{-1}$. 
$^c$ Kinetic luminosity in relativistic protons in units of $10^{48}$~erg~s$^{-1}$. $^d$ Partition fractions defined as 
$\epsilon_{ij} \equiv L_i/L_j$. $^e$ Minimum allowed variability time scale in hours.
}}
\end{table}
\end{rotate}

\subsection{\label{had_fits}Hadronic Model Fits}

\begin{figure}
\bigskip
\bigskip
\begin{center}
\vbox{\hbox{\includegraphics[width=6.8cm]{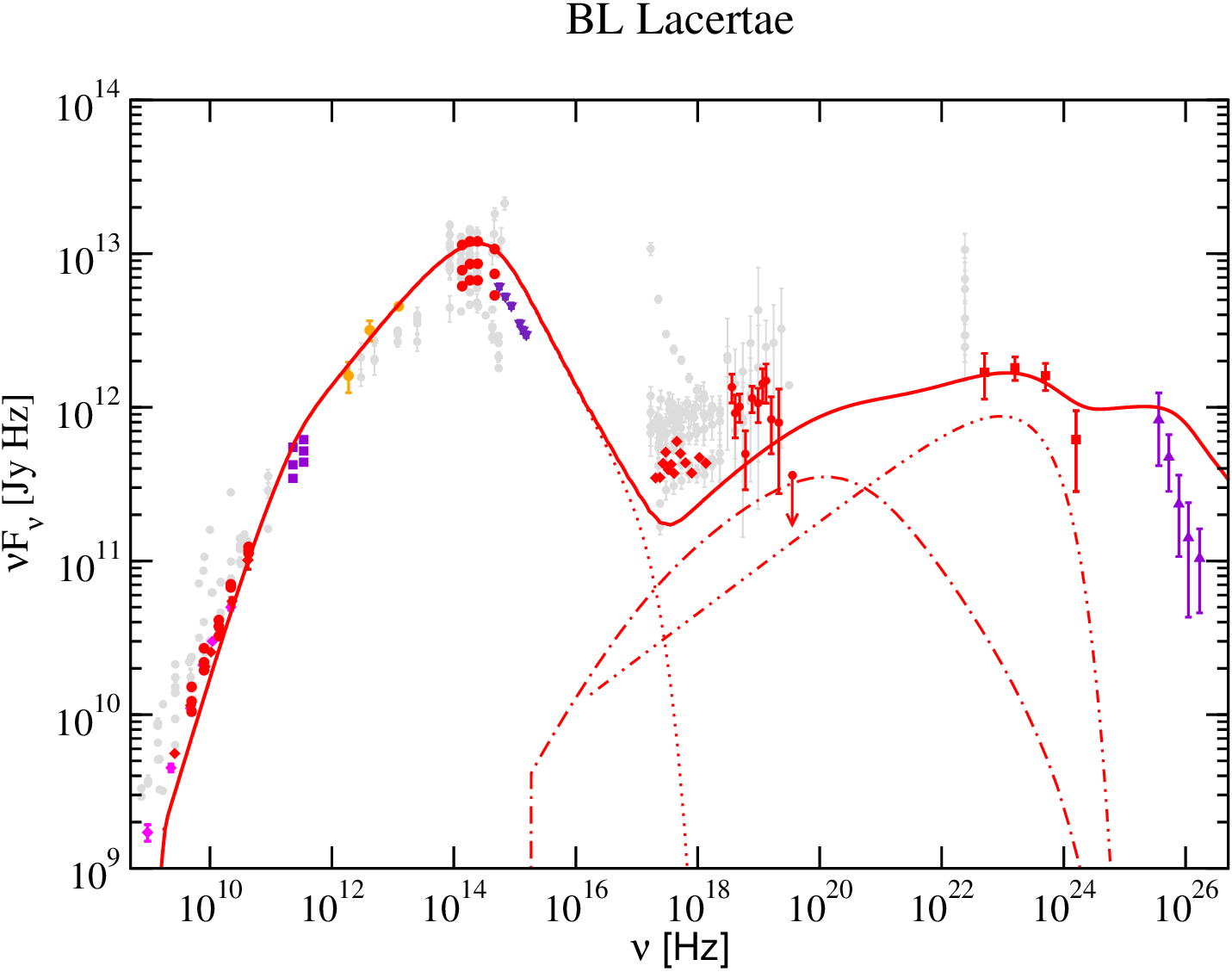}
            \hskip 1cm
            \includegraphics[width=6.8cm]{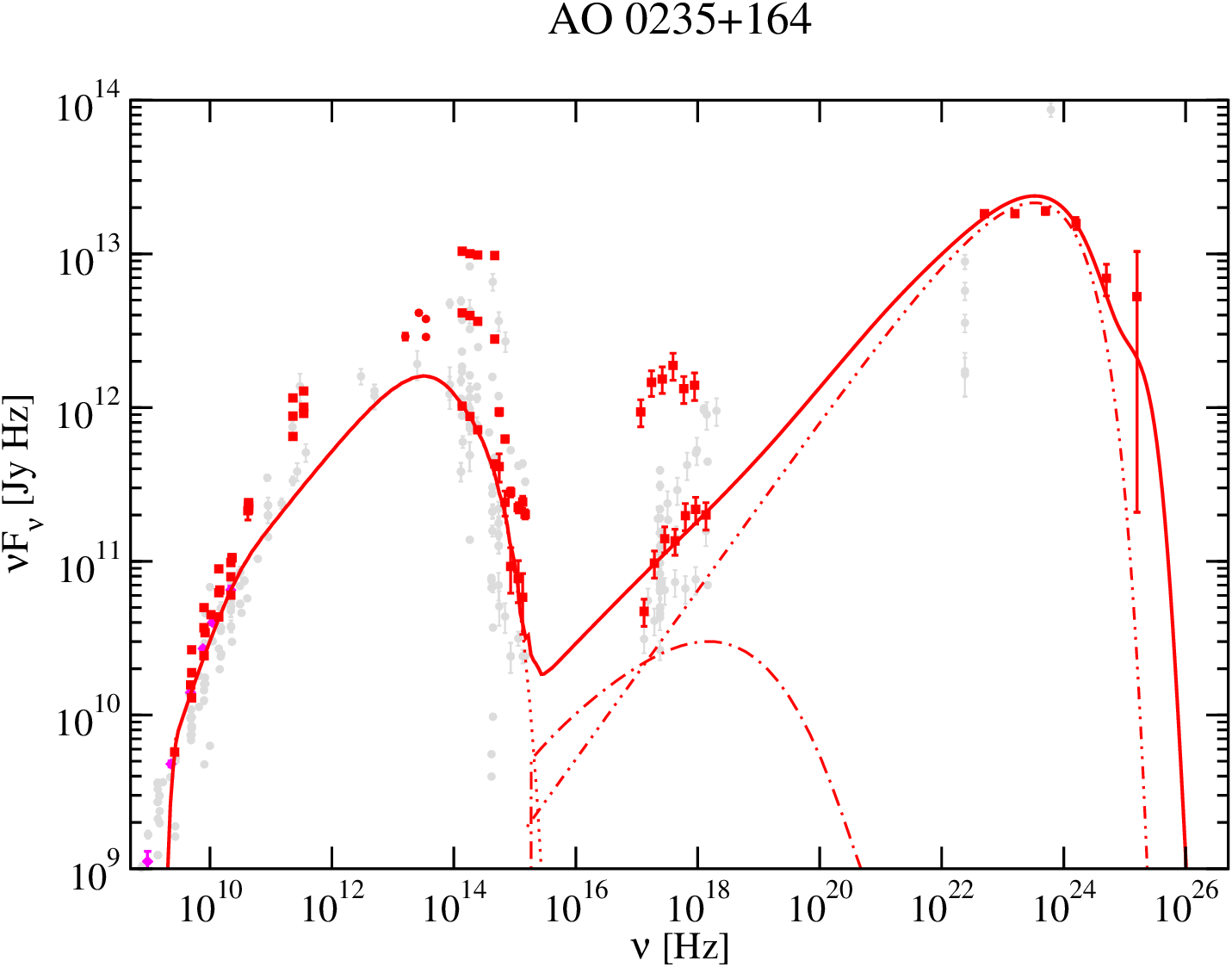}
            }
      \bigskip
      \hbox{\includegraphics[width=6.8cm]{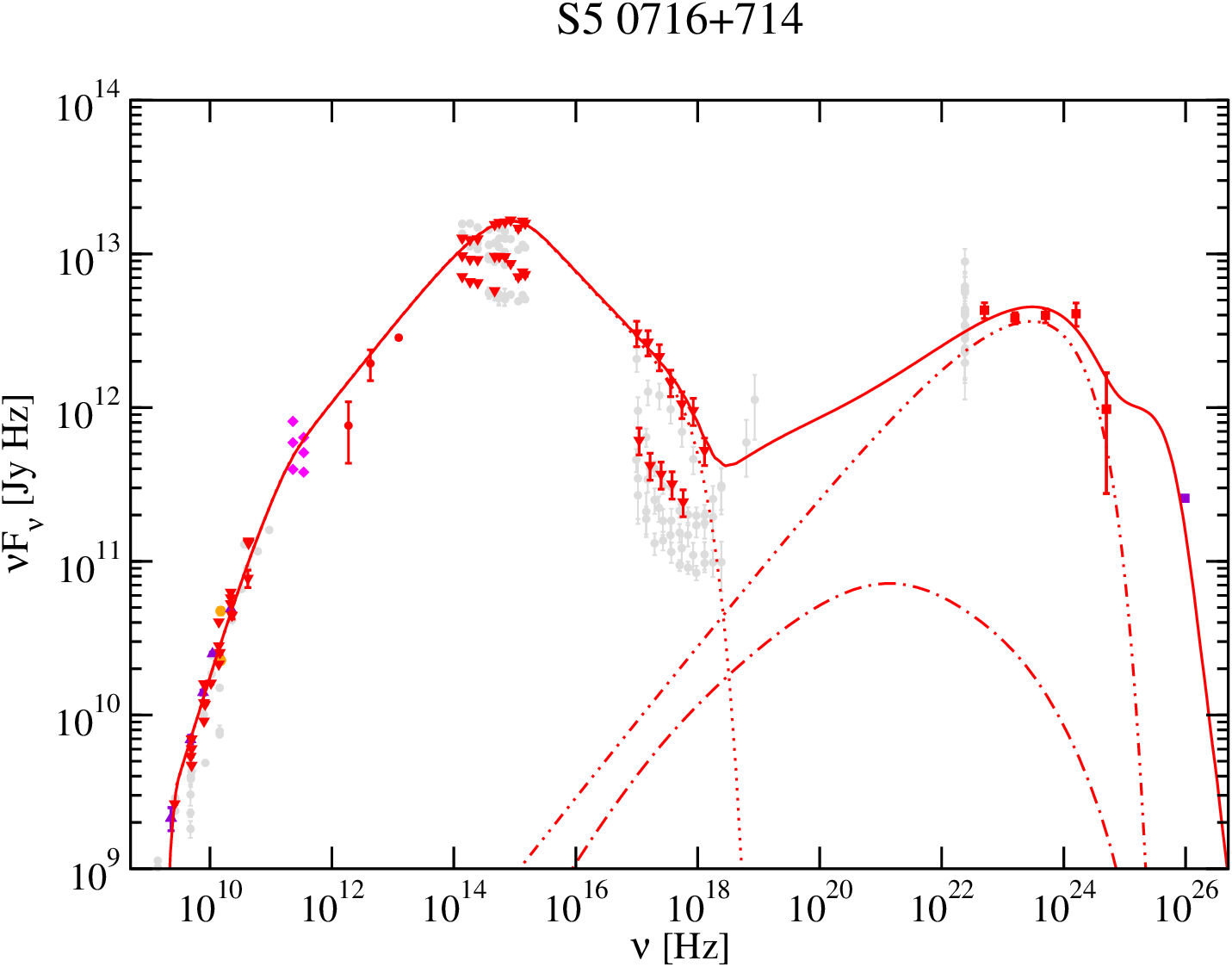}
            \hskip 1cm
            \includegraphics[width=6.8cm]{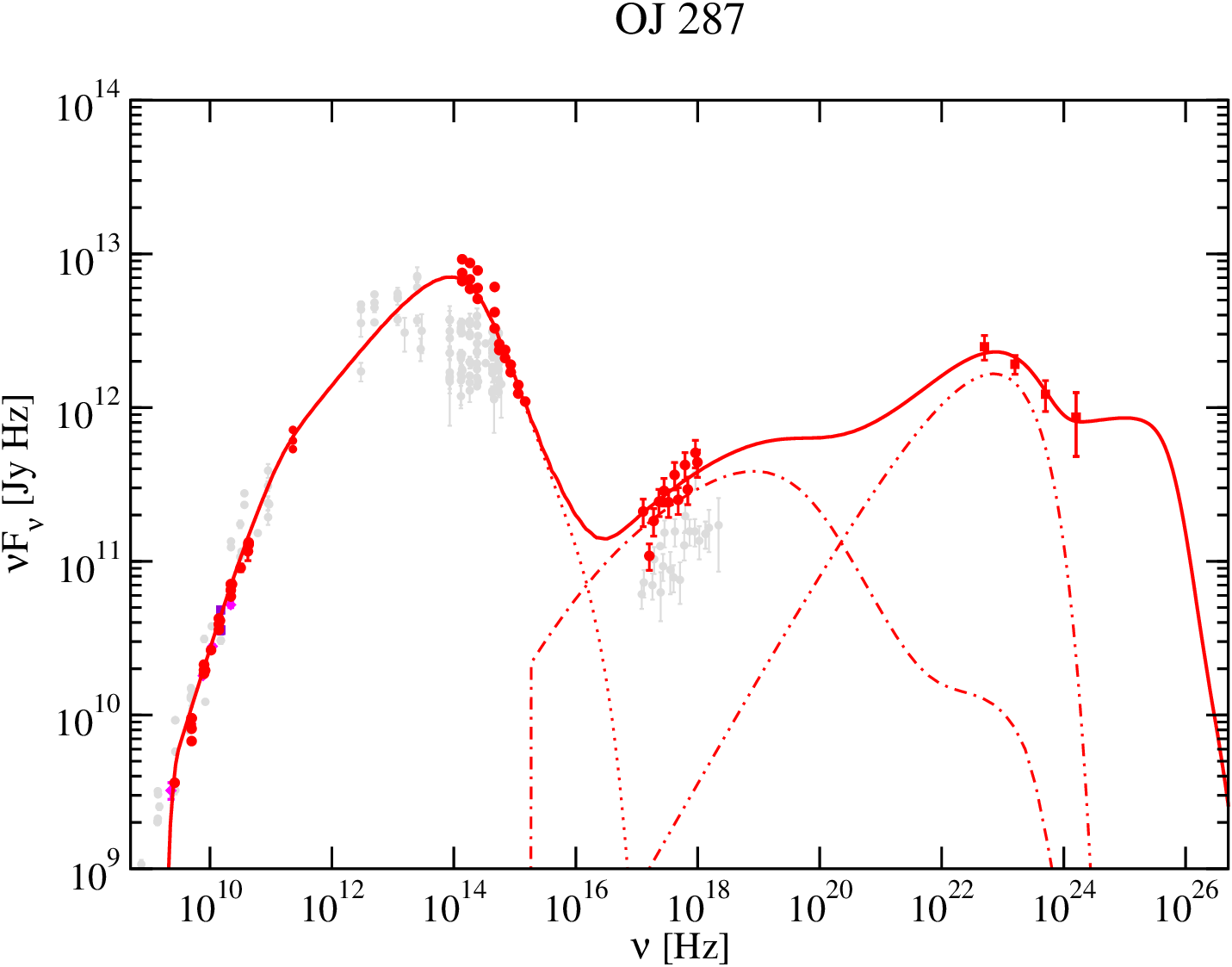}
            }
      }
\end{center}
\caption{\label{hadronic_LBL}Hadronic model fits to the 4 LBLs in
our sample. See table \ref{hfitpars} for parameters. }
\end{figure}

\begin{figure}
\bigskip
\bigskip
\begin{center}
\hbox{\includegraphics[width=6.8cm]{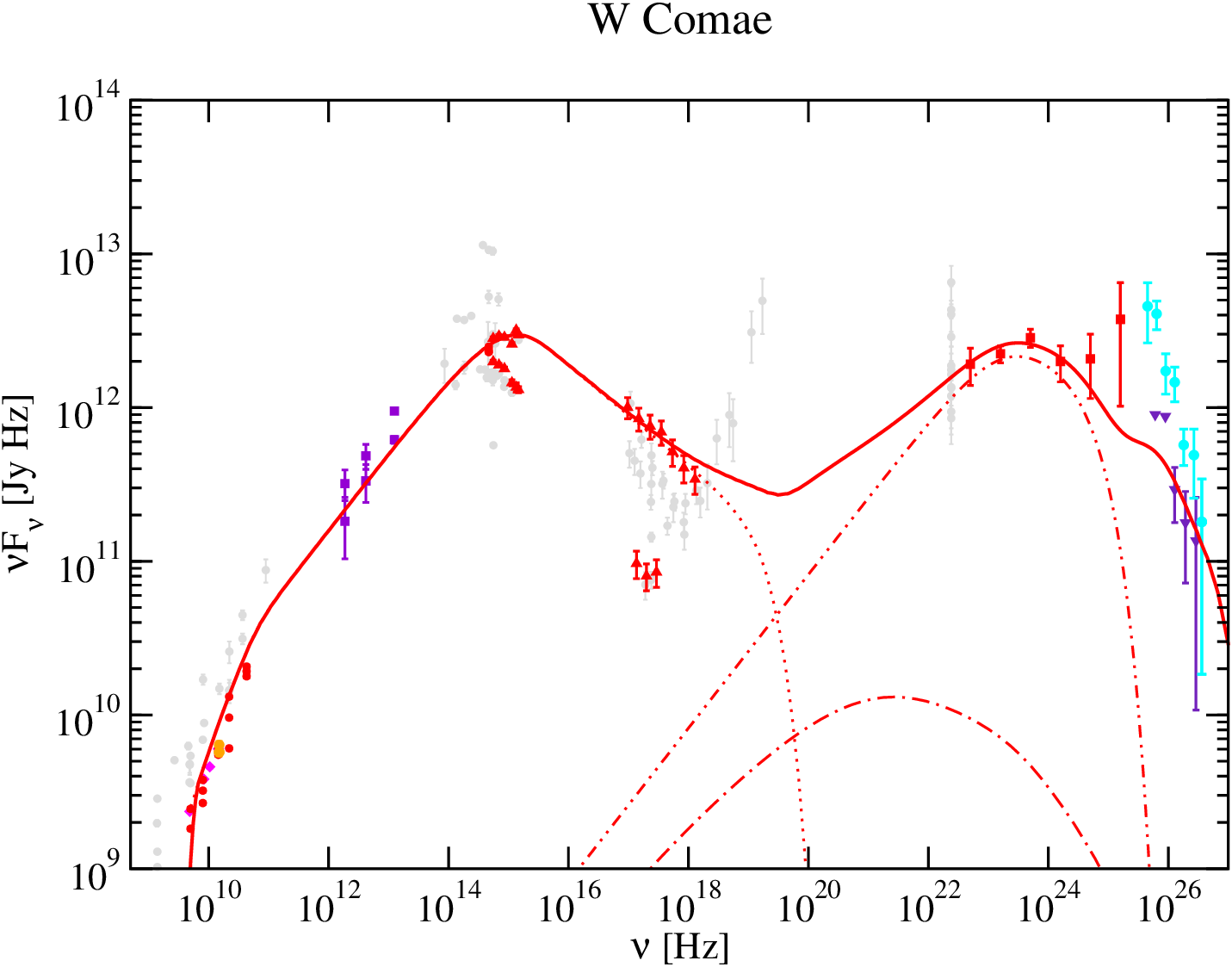}
      \hskip 1cm
      \includegraphics[width=6.8cm]{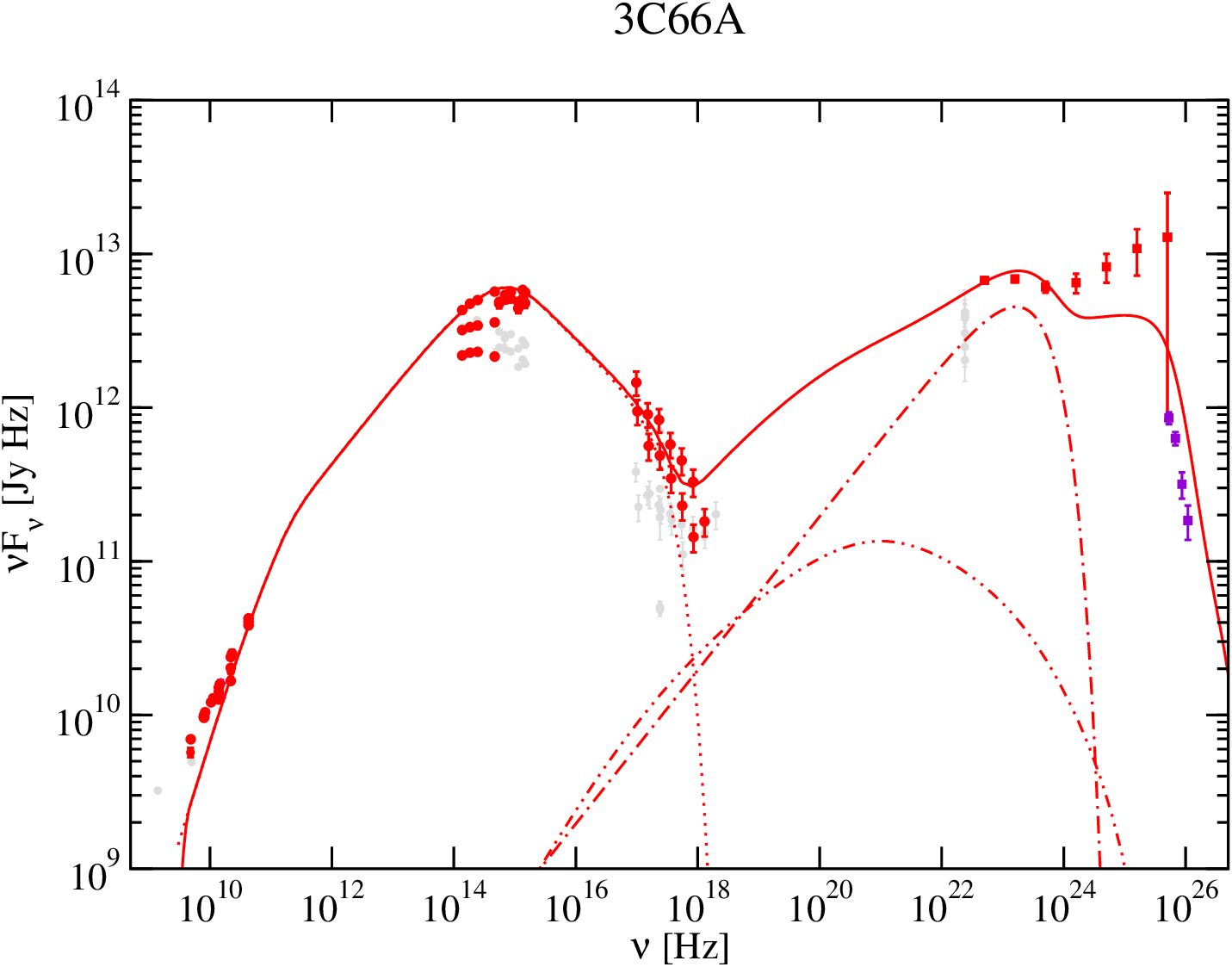}
      }
\end{center}
\caption{\label{hadronic_IBL}Hadronic model fits to the 2 IBLs in
our sample. See table \ref{hfitpars} for parameters. }
\end{figure}

Overall, we find that the SEDs of BL Lac objects, both IBLs (Fig.~\ref{hadronic_IBL}) and 
LBLs (Fig.~\ref{hadronic_LBL}), can be well represented with our model. The typically low 
ratio of $\gamma$-ray to X-ray flux and the hard $\gamma$-ray spectra observed in
BL Lac objects (as compared to FSRQs) is naturally obtained by photo-pion
induced cascade emission with substantial contributions from proton-synchrotron radiation.
In some cases, an additional contribution from the leptonic SSC emission aids in 
producing the observed X-ray flux. However, even though the addition of photopion
induced cascade emission to a proton synchrotron component is, in principle, able to
reproduce concave $\gamma$-ray spectra, the steep upturn of the Fermi-LAT spectrum of 3C~66A
and W~Comae at GeV energies is still not well represented also by this model, and may 
indicate the need for an additional emission component (possibly from muon/pion synchrotron 
radiation, with consequences for the values of the fit parameters). We note that the 
required power in relativistic protons, $L_p$, is very large, in the range $\sim 10^{47}$ --
$10^{49}$erg s$^{-1}$, which is significantly higher than the observed radiative luminosities 
of these objects.

\begin{figure}
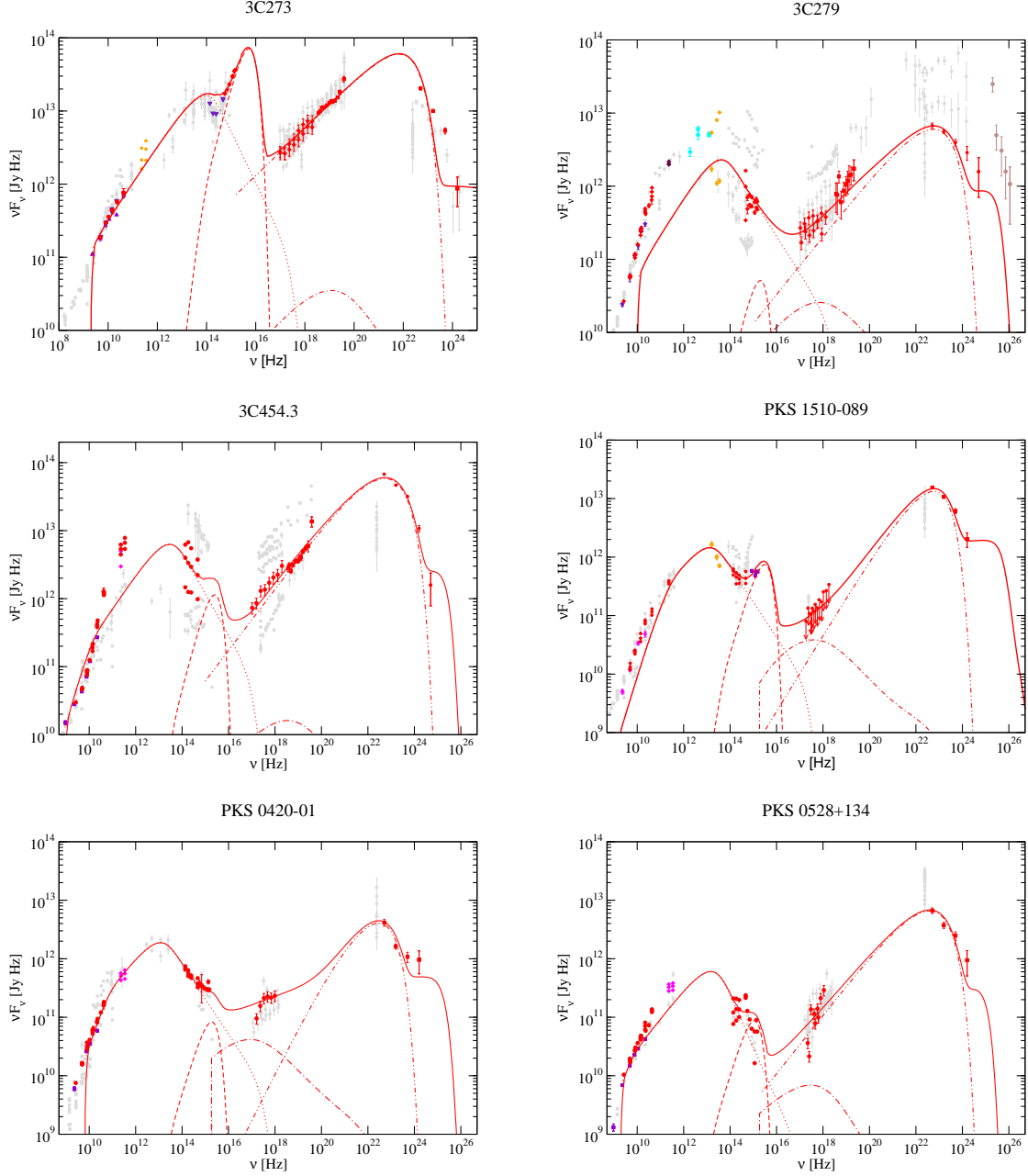

\bigskip
\bigskip
\begin{center}
\vbox{\hbox{\includegraphics[width=6.8cm]{3C273had.eps}
            \hskip 1cm
            \includegraphics[width=6.8cm]{3C279had.eps}
            }
      \bigskip
      \hbox{\includegraphics[width=6.8cm]{3C454.3had.eps}
            \hskip 1cm
            \includegraphics[width=6.8cm]{PKS1510had.eps}
            }
      \bigskip
      \hbox{\includegraphics[width=6.8cm]{PKS0420had.eps}
            \hskip 1cm
            \includegraphics[width=6.8cm]{PKS0528had2.eps}
            }
      }
\end{center}
\caption{\label{hadronic_FSRQ}Hadronic model fits to the 6 FSRQs in
our sample. See table \ref{hfitpars} for parameters. Dotted = electron-synchrotron;
dashed = accretion disk; dot-dashed = SSC; dot-dot-dashed = proton-synchrotron. }
\end{figure}

Fits within the framework of the hadronic model to the SEDs of the FSRQs in our 
sample are shown in Fig.~\ref{hadronic_FSRQ}. Although the hadronic model appears 
to satisfactorally fit most of the snap-shot SEDs, the typically observed spectral 
break at GeV energies seems problematic in the cases of 3C~273 and 3C~279.

Also here, the $\gamma$-ray component is dominated by proton synchrotron radiation, 
with some contribution from proton initiated pair cascade emission above $\sim 10$~GeV. 
The too steep decline of the proton synchrotron emission at a few GeV may indicate 
too steep a decline of the radiating proton distribution in the cutoff region 
(here considered a hard-cutoff), and hence possibly an impact of the acceleration 
mechanism in this region \citep[e.g.,][]{protheroe04,aharonian00}.
The observed hard X-ray to soft $\gamma$-ray spectral 
shapes can be fitted if proton synchrotron emission is the dominant radiation channel
(in order not to over-predict the X-ray flux due to reprocessing of UHE $\gamma$-ray 
emission in cascades). This indicates the presence of high magnetic field strengths in 
the emission region.

The hadronic FSRQ fits also require extreme jet powers, in some cases exceeding 
$10^{49}$erg s$^{-1}$ in relativistic protons.

In summary, we find that the hadronic model presented here provides 
satisfactory fits to most of the bright blazar SEDs of our sample.
However, the declining (in $\nu F_{\nu}$) broken power-law shape 
observed in many {\it Fermi}-LAT spectra of FSRQs causes problems 
for adequate fits in two cases within the limits of this model.
Fits to most objects have been achieved with magnetic fields 
in the range $B \sim 10$ -- 30~G. All fits required proton acceleration
to energies of $E_p \gtrsim 10^{17-18}$~eV. Nearly all model fits used 
a strong cascade component to aid modeling the high energy GeV data. 
This leads to the requirement of a disproportionately large proton 
luminosity as compared to the magnetic field luminosity. As a consequence 
in all hadronic model fit parameter sets presented here the total jet 
power is dominated by protons. Alternatively, hadronic models that take 
charged $\pi/\mu$ synchrotron radiation into account \citep[e.g.,][]{muecke03}
could lower the strength of the pair casade component, thereby reducing 
the proton contribution to the jet power at the expense of magnetic 
field power. In order to produce the observed (electron-synchrotron) 
IR -- UV emission, the model requires lower characteristic electron 
energies for FSRQs ($\gamma_{e,1} \sim 100$) than for BL~Lac objects 
($\gamma_{e,1} \lesssim 10^3$), along with harder electron spectra in 
BL~Lac objects. All our fits were achieved with characteristic Doppler 
factors of $D \sim 10$ -- 30.

\section{\label{summary}Summary and Conclusions}

In this paper, we have described the development of new implementations
of stationary, single-zone leptonic and hadronic models. Our leptonic
model allows for arbitrary external photon sources, and solves self-consistently
for an equilibrium between relativistic particle acceleration (injection),
radiative cooling, and escape. Our hadronic model is based on the
\cite{ka08} templates for the final products of photo-pion production,
and uses a new, semi-analytical method for calculating the output from
ultra-high-energy induced pair cascades.

We have used both leptonic and hadronic models to fit the contemporaneous
SEDs of 12 {\it Fermi}-LAT deteced blazars with good multiwavelength
coverage and additional observational constraints on model parameters. 
We find that the SEDs of all types of blazars can be well represented
with leptonic models with parameters close to equipartition between the
magnetic field and relativistic electrons in the emission region. 
However, our leptonic model is unable to provide a good fit to the 
hard {\it Fermi}-LAT spectrum of AO~0135+164. The problem lies in
the mismatch between the very steep synchrotron (IR -- optical -- UV)
continuum, as opposed to the very hard $\gamma$-ray spectrum, and 
Klein-Nishina effects at the highest $\gamma$-ray energies. We
confirm that even intermediate BL~Lac objects are more appropriately
fit including an external radiation field as source for Compton scattering
to produce the observed $\gamma$-ray emission. FSRQs are characterized
by systematically more powerful jets, but lower bulk Lorentz factor and
softer electron spectra than BL~Lac objects. If charge neutrality in
blazar jets is provided by cold protons (rather then relativistic
positrons), our fits indicate that those protons are dominating the
kinetic power carried by the jets by about an order of magnitude.

The hadronic model presented here has difficulty describing the GeV-break 
in the SEDs of two FSRQs, but provides appropriate fits for all other blazars 
in our sample. However, the fits require very large powers in relativistic 
protons, of $L_P \sim 10^{47}$ -- $10^{49}$~erg~s$^{-1}$, in most cases dominating
the total power in the jet.

\acknowledgements{We thank Paolo Giommi for sending us the SED data for
our modeling, and the anonymous referee for a helpful and constructive
report. MB acknowledges support by the South African Department of 
Science and Technology through the National Research Foundation under
NRF SARChI Chair grant no. 64789. This work was supported by NASA through 
Astrophysics Theory Program Award NNX10AC79G and Fermi Guest Investigator
Grant NNX09AT82G. AR acknowledges support by Marie Curie IRG grant 248037 
within the FP7 Program.}

\end{document}